\documentclass[aps,prb,showpacs,amsmath,amssymb,superscriptaddress,a4paper,twocolumn]{revtex4-1}
\usepackage[utf8]{inputenc}
\usepackage{amsmath}
\usepackage{amsfonts}
\usepackage{amssymb}
\usepackage{graphicx}
\usepackage{textgreek}
\usepackage{miller}
\usepackage{physics}
\usepackage[version=4]{mhchem}
\usepackage{xcolor}

\begin{document}


\author{Dags Olsteins}
\affiliation{Center For Quantum Devices, Niels Bohr Institute, University of Copenhagen,\\ 2100 Copenhagen, Denmark}

\author{Gunjan Nagda}
\affiliation{Center For Quantum Devices, Niels Bohr Institute, University of Copenhagen,\\ 2100 Copenhagen, Denmark}

\author{Damon J. Carrad}
\affiliation{Department of Energy Conversion and Storage, Technical University of Denmark, 2800 Kgs.Lyngby, Denmark}

\author{Daria V. Beznasiuk}
\affiliation{Department of Energy Conversion and Storage, Technical University of Denmark, 2800 Kgs.Lyngby, Denmark}

\author{Christian E. N. Petersen}
\affiliation{Department of Energy Conversion and Storage, Technical University of Denmark, 2800 Kgs.Lyngby, Denmark}

\author{Sara Martí-Sánchez}
\affiliation{Catalan Institute of Nanoscience and Nanotechnology (ICN2), CSIC and BIST, Campus UAB, Bellaterra, Barcelona, Catalonia, Spain}

\author{Jordi Arbiol}
\affiliation{Catalan Institute of Nanoscience and Nanotechnology (ICN2), CSIC and BIST, Campus UAB, Bellaterra, Barcelona, Catalonia, Spain}
\affiliation{ICREA, Passeig de Lluís Companys 23, 08010 Barcelona, Catalonia, Spain}

\author{Thomas Sand Jespersen}
\affiliation{Center For Quantum Devices, Niels Bohr Institute, University of Copenhagen,\\ 2100 Copenhagen, Denmark}
\affiliation{Department of Energy Conversion and Storage, Technical University of Denmark, 2800 Kgs.Lyngby, Denmark}
\email{tsaje@dtu.dk}

\title{Cryogenic Multiplexing with Bottom-Up Nanowires}

\begin{abstract}
Bottom-up grown nanomaterials play an integral role in the development of quantum technologies. Among these, semiconductor nanowires (NWs) are widely used in proof-of-principle experiments, however, difficulties in parallel processing of conventionally-grown NWs makes scalability unfeasible. Here, we harness selective area growth (SAG) to remove this road-block. We demonstrate large scale integrated SAG NW circuits consisting of 512 channel multiplexer/demultiplexer pairs, incorporating thousands of interconnected SAG NWs operating under deep cryogenic conditions. Multiplexers enable a range of new strategies in quantum device research and scaling by increase the device count while limiting the number of connections between room-temperature control electronics and the cryogenic samples. As an example of this potential we perform a statistical characterization of large arrays of identical SAG quantum dots thus establishing the feasibility of applying cross-bar gating strategies for efficient scaling of future SAG quantum circuits. 
\end{abstract}

\maketitle


\begin{figure*}
  \centering
    \includegraphics[width = 16 cm,page=1]{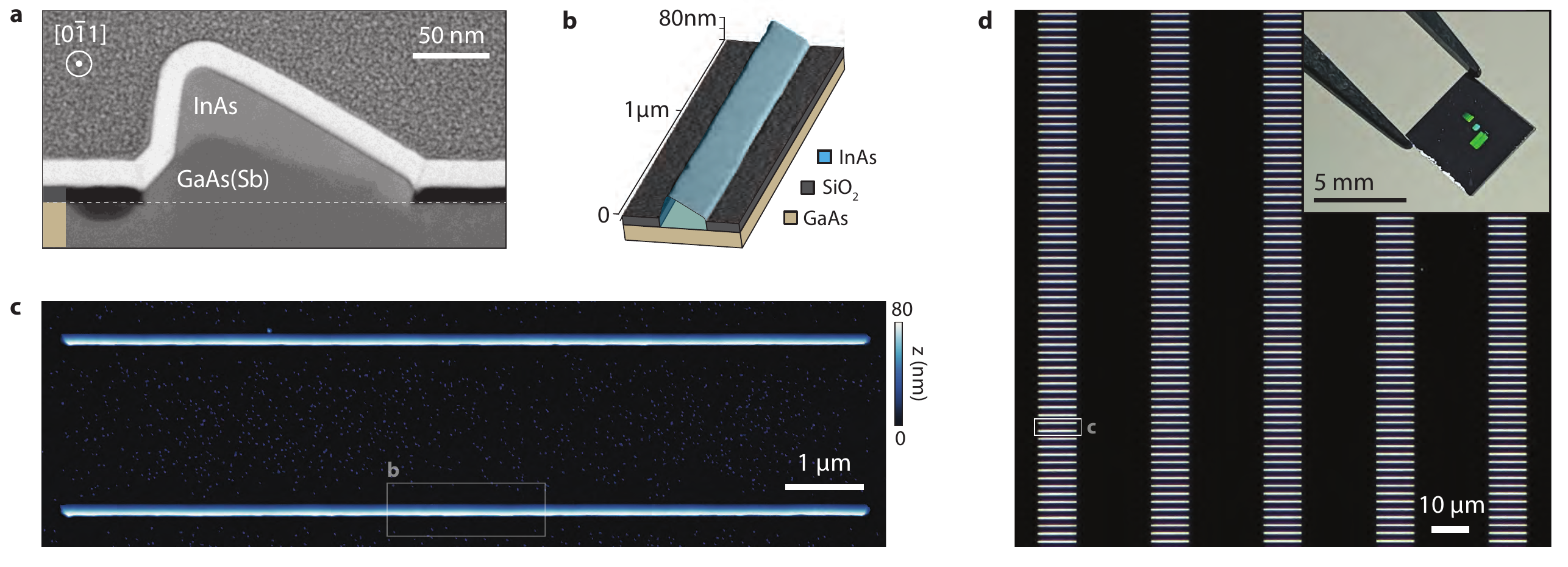} 
    \caption{ \textbf{a} Cross-sectional HAADF STEM micrograph of a SAG NW, showing the InAs conducting channel atop the GaAs substrate and GaAs(Sb) MBE-grown buffer. The NW exhibits an asymmetric triangular shape imposed by the \hkl(311) substrate symmetry. The shape is not important for the present study which could equally well have been based on NWs grown on, e.g., \hkl(111) or \hkl(100) substrates with higher symmetry. \textbf{b} Combined schematic/3D atomic force microscope (AFM) micrograph of a $2 \, \mu \mathrm m$ section of a single NW. \textbf{c} AFM micrograph of two SAG NWs. \textbf{d} Optical dark-field microscope image of a section of an InAs SAG NW array. Each NW is $\sim 150 \, \mathrm{nm}$ wide and $10 \, \mu \mathrm m$ long and individual NWs are spaced by $20 \times 2 \, \mu \mathrm m$. Inset: Photograph of an as-grown sample. The large NW arrays are visible in green due to diffraction of light.
    } 
    \label{Fig1}
  \hfill
\end{figure*}

Quantum electronics is rapidly maturing towards large scale integrated (LSI) circuits incorporating a multitude of interacting quantum devices. There is therefore an onus on potential quantum materials candidates to exhibit both high-precision reproducibility and scalability potential. Semiconductor nanowires (NWs) constitute an important platform for quantum electronics, since the electronic confinement intrinsic to the structure simplifies fabrication of complex devices\cite{mourik:2012,hu:2007,hofstetter:2009,fasth2005tunable}, the flexibility of contact materials enables hybridization with important quantum materials such as superconductors \cite{doh:2005, xiang:2006, hofstetter:2009, mourik:2012,larsen:2015,krogstrup:2015,hays:2018}, and an increased capacity for strain relaxation over bulk materials enables exploration of exotic heterostructures\cite{hu:2007,xiang:2006,krogstrup:2015,SvenssonNL15}. Conventional NWs grown perpendicular to the substrate have been highly successful for high-performance electronics, simple complementary circuits, and in fundamental mesoscopic physics\cite{fasth2005tunable,hu:2007,doh:2005, xiang:2006, hofstetter:2009, mourik:2012,larsen:2015,krogstrup:2015,hays:2018,tomioka:2012,SvenssonNL15}. However, the difficulty in fabricating individual vertical devices, and the insufficient precision and yield of techniques for transferring NWs to the planar geometry compatible with standard semiconductor processing\cite{freer:2010a,yao:2013,li:2008} has thus far inhibited development of LSI NW circuits. A promising alternative is the bottom-up growth of in-plane semiconductor NWs directly on a suitable device substrate using selective area growth\cite{Wang2019Jun,op2020plane, Raya2020Jan, Bollani2020Jan} (SAG). In SAG, the positions and dimensions of NWs are controlled by lithographically defining openings in a dielectric mask, enabling the controlled growth of large-scale networks and NW arrays\cite{Aseev2019Jan,Krizek2018Sep, op2020plane}. While proof-of-principle single NW devices, e.g., field effect transistors (FETs) \cite{beznasyuk:2022,Friedl2018Apr}, Hall crosses\cite{Lee2019Aug}, quantum interferometers\cite{vaitiekenas2018selective,Krizek2018Sep}, hybrid superconducting devices\cite{hertel:2022, vaitiekenas2018selective}, and quantum dots\cite{tenkate:2022} have been reported, scalability towards integrated quantum circuits -- a central motivation behind the development of SAG and similar bottom-up grown planar nanostructures\cite{schmid:2015, gooth:2017} -- has not been addressed. 

Here we make the first demonstration of LSI circuits based on SAG. Starting from large arrays of thousands of SAG NWs we fabricate multiplexer (MUX) circuits which operate at the deep cryogenic conditions relevant for quantum electronics. Cryogenic multiplexers are key ingredients towards scaling of quantum electronics\cite{pauka:2020,paquelet2020multiplexed,smith:2014,smith:2020} as the number of addressable devices scales exponentially -- rather than linearly -- with the number of connecting control lines. This is crucial for reducing heat load from wiring between the cryogenic sample and room-temperature, and integrated MUX circuits allow highly dense packing of devices utilizing chip-area conventionally required for bonding and routing. Our setup allows us to address and measure 512 individual SAG quantum devices using only 37 control lines. Our architecture also includes de-multiplexers (d-MUX) connected back-to-back with the corresponding MUX, enabling us to unambiguously confirm the functionality of the circuit, to identify faulty operation among the thousands of NWFETs, and self-correct against most failure modes.

Introducing on-chip multiplexing to bottom-up grown nanostructures enables new strategies in quantum electronics research, such as automated searches through large ensembles of devices for rare or exotic phenomena, and systematic, statistically significant exploration of the correlation between device performance and e.g., materials properties or device geometry. To demonstrate the potential of the latter we perform a statistical characterization of device reproducibility within a large array of nominally identical SAG NW quantum dots (QDs). QD arrays are promising candidates for implementing quantum computation and simulation\cite{loss1998quantum, tomioka2020rational, vandersypen2019semiconductor}, and quantifying device-to-device reproducibility -- enabled by the MUX circuit -- is crucial for the successful development of crossbar gate architectures which constitute an important strategy for limiting gate counts in realistic large scale implementations\cite{li2018crossbar,Borsoi:2022}. We find that all QDs of the array can be concurrently tuned to Coulomb Blockade using only three shared crossbar gates further confirming the potential SAG as a scalable platform for quantum devices,

\begin{figure}
  \centering
    \includegraphics[width = \linewidth,page=1]{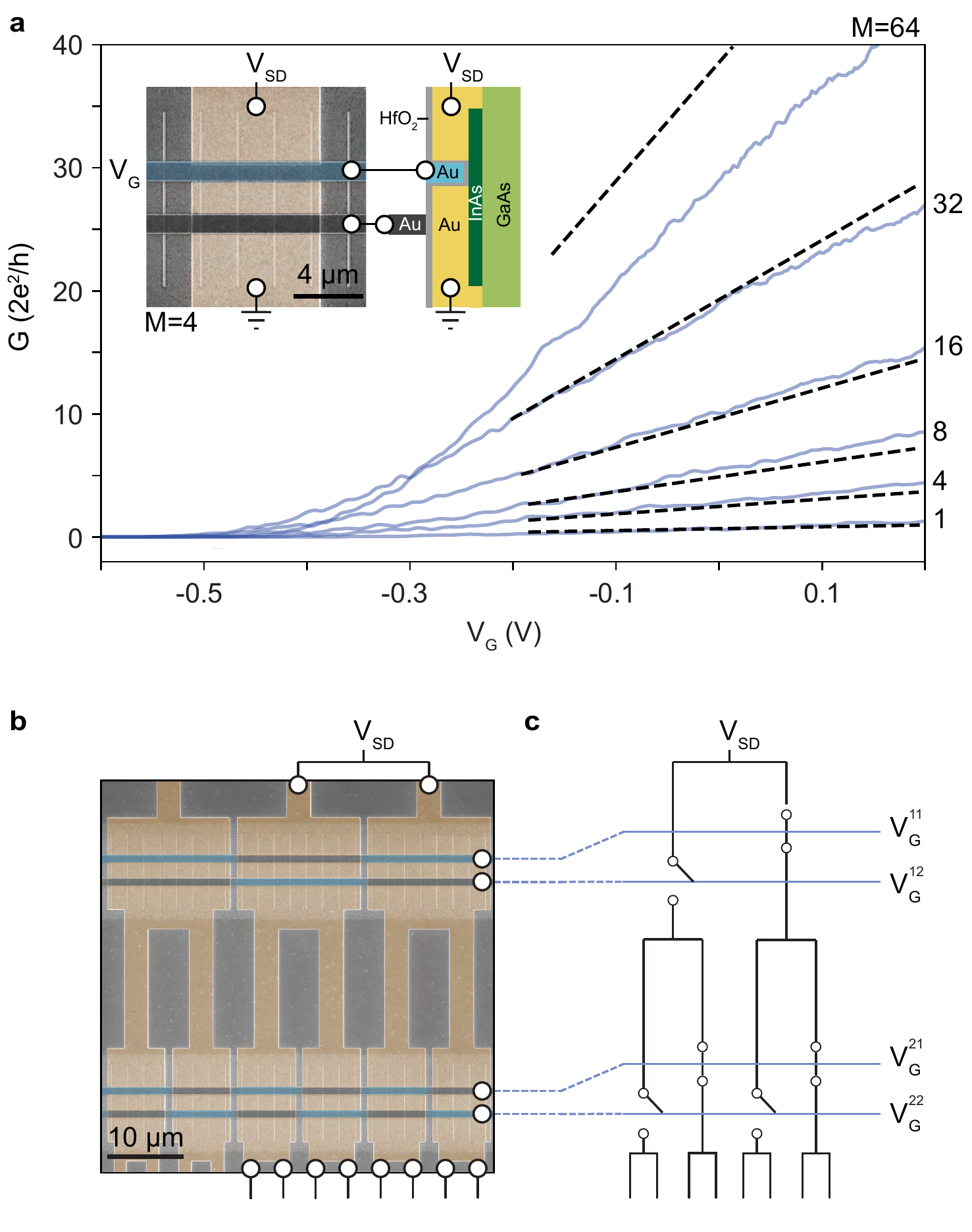} 
    \caption{\textbf{a} Conductance, $G$, as function of gate voltage V\textsubscript{G} for NWFETs based on 1, 4, 8, 16, 32, and 64 SAG NWs in parallel. Inset shows an SEM micrograph of a NWFET based on 4 SAG NWs, Ti/Au Ohmic contacts in gold, Ti/Au gate in blue. The gate in dark gray is screened by the underlying ohmic contact as illustrated by the cross-section schematic. \textbf{b} SEM micrograph of NWFETs arranged in a MUX circuit. Gates are screened in alternating elements as indicated by the blue/gray false coloring. \textbf{c} A schematic representation of the circuit in b.} 
    \label{Fig2}
  \hfill
\end{figure}

\section*{Material and electrical properties}
Our circuits are based on \hkl[0-11] oriented InAs SAG NWs grown using molecular beam epitaxy (MBE) on GaAs \hkl(311)A substrates. See Methods and Supplementary Section S1 for details. Figure~\ref{Fig1}a shows a cross-sectional high-angle annular dark field scanning transmission electron microscope (HAADF STEM) micrograph of a single NW, and Fig.~\ref{Fig1}b shows a combined schematic and atomic force microscopy (AFM) micrograph of a NW section. The conducting InAs channel sits atop an insulating GaAs substrate and GaAs(Sb) buffer\cite{Krizek2018Sep}. The NWs are terminated with \hkl{111}A facets as a consequence of the \hkl(311)A substrate symmetry, producing the asymmetric cross-section. Detailed structural analysis is presented in Supplementary Section S2. Figures~\ref{Fig1}c and d illustrate the capacity for scale-up inherent to SAG, through an AFM micrograph and dark-field optical microscope micrograph of representative sections of a $512 \times 16$ array of nominally identical $10~\mu$m-long NWs. The inset to Fig.~\ref{Fig1}d shows a photograph of a cleaved $5 \times 5\, \mathrm{mm}$ piece of the growth wafer containing $\sim 18000$ SAG NWs, $9216$ of which were used for device fabrication; the diffraction from the large arrays is visible.

\begin{figure*}
  \centering
    \includegraphics[width = \linewidth]{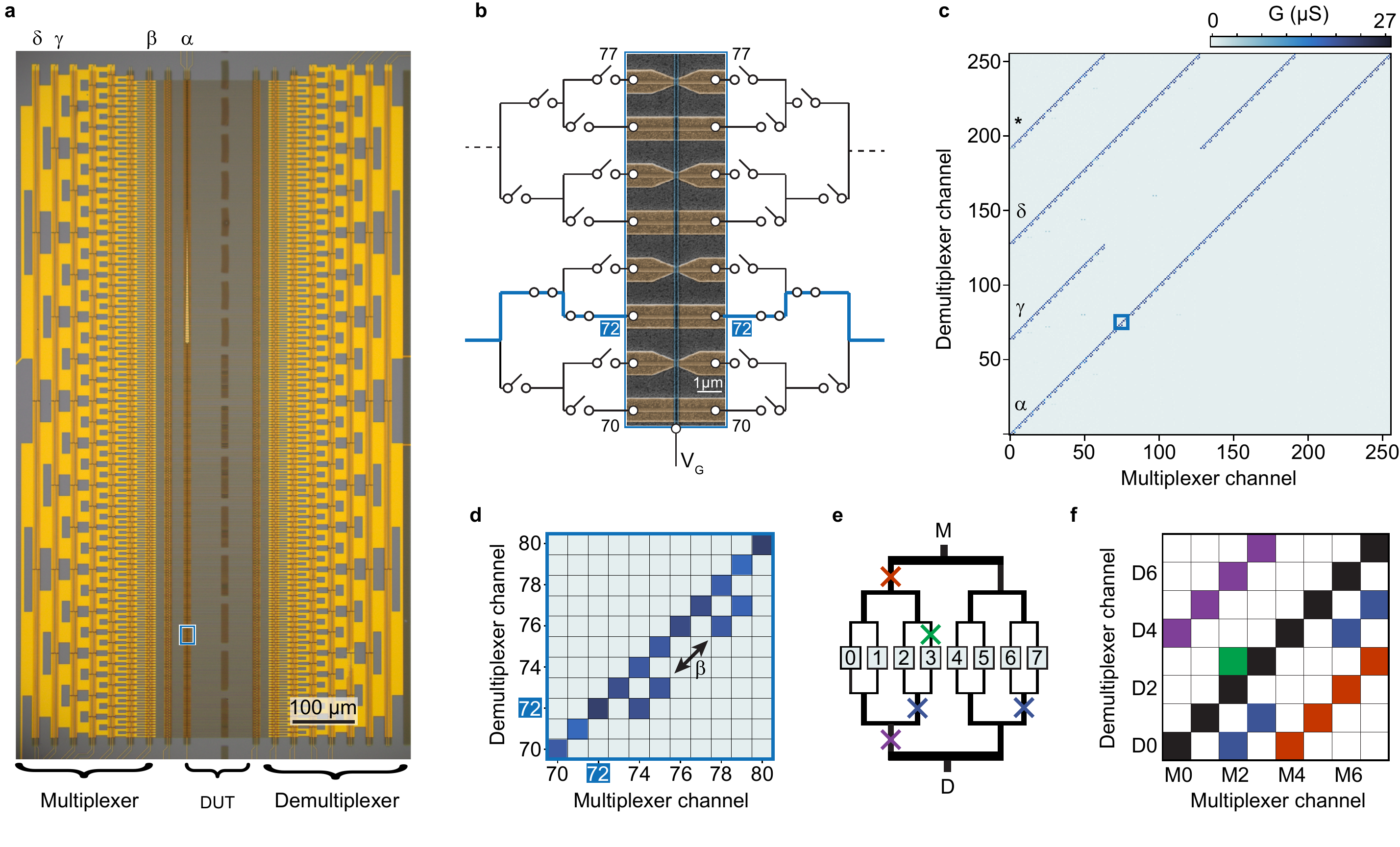}
    \caption{\textbf{a} Optical microscope image of MUX/d-MUX circuit based on InAs SAG NW arrays. Devices under test (DUT) are labeled $\alpha$ and labels $\beta, \gamma, \delta$ indicate FET gates failing to deplete and related to corresponding off-diagonal signals in panel c. For the device shown 500 of the 512 lines were connected at the DUT level. The number of DUT was doubled from $2^8 = 256$ by using two source/drain pairs. \textbf{b} SEM micrograph of the DUT area indicated in panel a, showing 8 devices connected to the MUX and d-MUX channels. Every second channel is shorted for use as a reference to obtain the series resistance of the adjacent channel. \textbf{c} Conductance matrix of all 65536 combinations of source and drain channels of the first 256 connections of the circuit. The color of each pixel corresponds to the measured conductance of the specific channel combination. The uninterrupted diagonal feature shows that that all of the DUT are addressable and conducting. \textbf{d} Expanded view of the conductance matrix within the blue square in \textbf{c}. \textbf{e} Schematic three-level MUX/d-MUX. The colored crosses mark a FET failing to deplete and panel \textbf{f} shows the corresponding signatures on the conductance matrix. When operated at the diagonal, the circuit is immune to such errors except symmetric pairs such as the red/purple. This case can, however, be identified in the conductance matrix as regions with symmetric off-diagonal finite conductance, and accounted for in subsequent measurements.
    } 
    \label{Fig3}
\end{figure*}

\begin{figure*}
  \centering
    \includegraphics[width = \linewidth]{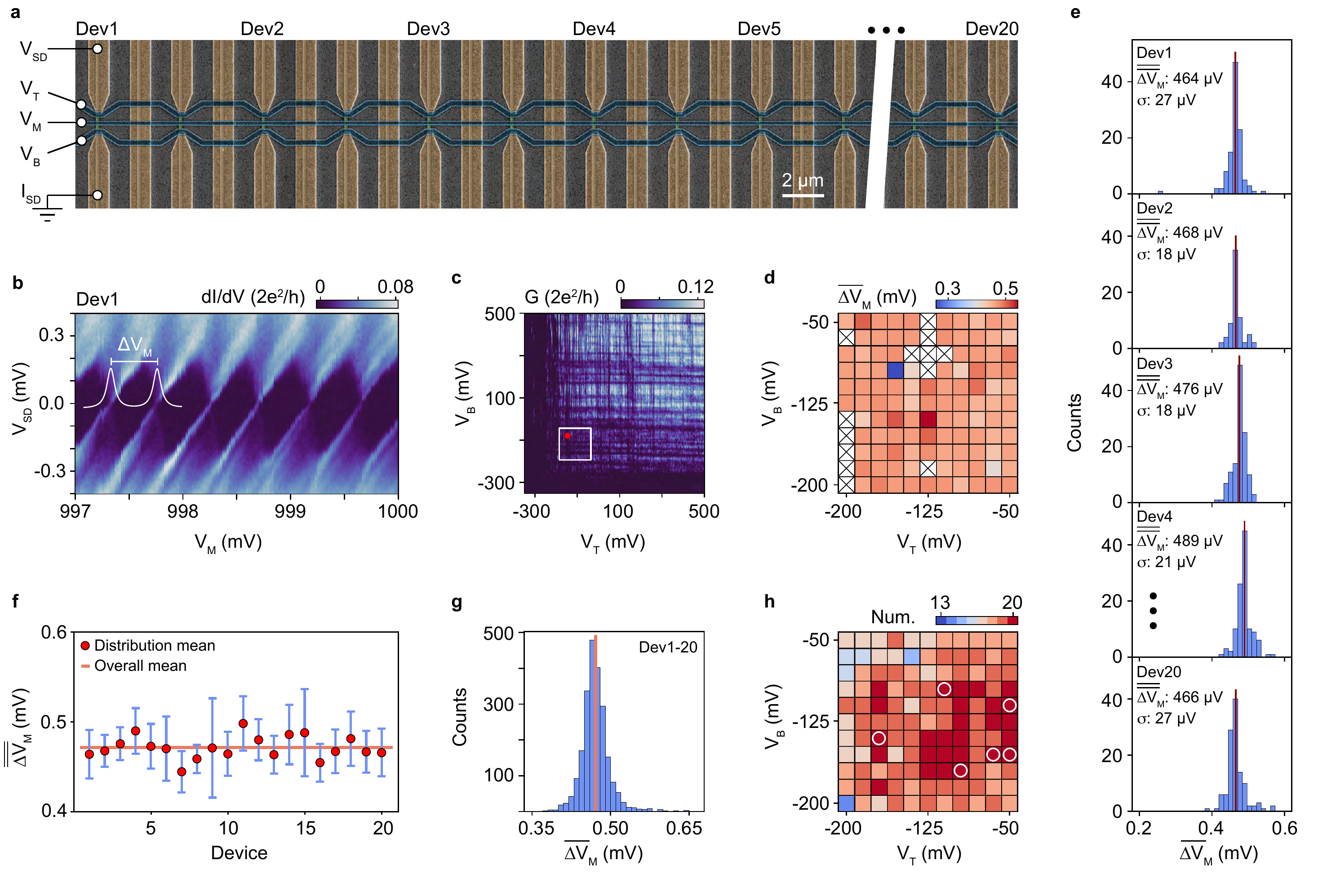}
    \caption{\textbf{a} SEM micrograph of an array of NW QD devices. Contacts (gold) are spaced by 900 nm, top/middle/bottom cross-bar gates (blue) are 300/200/300 nm wide and have a 150 nm spacing. The gates are shared between 50 devices of which 20 were analysed in detail. Each device is paired with a shorted line for reference. \textbf{b} Source-drain bias spectrum (differential conductance $dI/dV$ vs $V_\mathrm{sd}$ and $V_\mathrm{M}$) of Dev1 showing Coulomb diamonds. When sweeping $V_\mathrm M$ the barrier gates were compensated for a slight capacitive cross coupling (Supplementary Section S10). $V_\mathrm{T}$ and $V_\mathrm{B}$ starting values correspond to the red point in c. \textbf{c} Conductance $G$ vs $V_\mathrm{T}$ and $V_\mathrm{B}$ for Dev1. \textbf{d} Map showing $\overline{\Delta V_\mathrm{M}}$ of Dev1 for various $V_\mathrm{T}$ and $V_\mathrm{B}$ combinations within the white square in c. Crossed points have no identifiable Coulomb peaks. \textbf{e} Histograms of $\overline{\Delta V_\mathrm{M}}$ for Dev1-Dev4 and Dev20 in the same $V_\mathrm{T}$ and $V_\mathrm{B}$ range. \textbf{f} Distribution of $\overline{\Delta V_\mathrm{M}}$ for Dev1-Dev20. The red dots indicate $\overline{\overline{\Delta V_\mathrm{M}}}$ of each device and red dashed line shows $\overline{\Delta V_\mathrm{M}}$ across all devices. \textbf{g} Histogram showing overall distribution of $\overline{\Delta V_\mathrm{M}}$ for all devices. Red dashed line indicate the average value and corresponds to the line in \textbf{f}. \textbf{h} Number of devices with observable Coulomb peaks at each combination of $V_\mathrm{T}$ and $V_\mathrm{B}$. White circles show points where all 20 devices have $\overline{\Delta V_\mathrm{M}}$ within $2\sigma$ of the common average.} 
    \label{Fig4}
\end{figure*}

The Fig.~\ref{Fig2}a inset shows a scanning electron microscope (SEM) micrograph of a typical device along with a schematic cross section. The device includes $4$ SAG NWs connected in parallel by Ti/Au ohmic contacts and a $1 \, \mu \mathrm m$ gated segment (see Methods for fabrication details). Two gates are seen: one which acts on the exposed InAs NWs (blue), thereby controlling the conductivity, and one which is screened by the metal contact, and thus has no effect on the underlying NW (grey). This gate is, however, important for the MUX operation as discussed below. Figure~\ref{Fig2}a shows the conductance, $G$, as a function of gate voltage, $V_\mathrm G$, measured at a temperature of $T=20$~mK for 6 different devices with varying numbers of NWs, $M=1,4,16,32,64$ after subtraction of a constant series resistance $R_\mathrm S$ (see Methods). The devices act as normally-on, $n$-type FETs with identical threshold voltages and the dashed lines show a common fit to the relation $G = KM(V_\mathrm{G} - V_\mathrm{TH})$ with fixed parameters $K = 0.12\, \mathrm{mS/V}$ and $V_\mathrm{TH} = -0.4 \, \mathrm V$. Except for $M=64$ where $G$ is somewhat lower than expected, $G$ is proportional to $M$ as expected for equally contributing NWs and the linear scaling with $V_\mathrm G$ is typical for NW FETs. The deviation for $M=64$ may be due to a high sensitivity to the estimate of $R_\mathrm S$ when the device resistance is low (Methods). Importantly, Fig.~2a shows that SAG devices manufactured in parallel exhibit consistent $G$ vs $V_\mathrm G$, with reproducible, $M$-independent $V_\mathrm{TH}$, enabling the use of large-$M$ FETs as building blocks in LSI SAG circuits. 

\section*{SAG multiplexers}
We utilize the $V_\mathrm{TH}$ reproducibility to operate the circuit shown in Fig.~\ref{Fig2}b. SAG FETs are connected in a hierarchical MUX structure, with each level consisting of devices fabricated on different rows of the SAG NW array in Fig.~\ref{Fig1}d. Each gate spans the respective row, and the positions of the gated segment alternate such that for each NW FET, one gate (blue) tunes the carrier density of the NWs while the other is screened (grey). Input signals are thus directed through the MUX as illustrated in the schematic in Fig.~\ref{Fig2}c. With this design, each additional level doubles the number of outputs such that a $n$ level MUX has $2^n$ outputs and requires $2n$ gates for operation.

Figure \ref{Fig3}a shows an optical micrographs of an 8-level MUX circuit connected back-to-back to a corresponding 8-level d-MUX. The circuit has a footprint of $\sim 0.6 \times 1.1 \, \mathrm{mm}$ and incorporates 8192 individual SAG NWs in the form of 1996 interconnected FETs. Combining the 32 gate lines with two separate source-drain pairs enables individual addressing of any of 512 devices under test (DUT) located in the gap between the MUX/d-MUX units. Where the DUT themselves consist of FETs with a single common gate, 37 control lines thereby enable experiments on 512 devices. In our case, the DUT in Fig.~\ref{Fig3}a consist of SAG devices with different functionalities and properties. For example, the SEM micrograph in Fig.~\ref{Fig3}b shows DUT devices \#70 - 77. Odd-numbered devices \#71,73,75,77 are SAG NWFETs with a contact separation of 100 nm and a common top gate. The even-numbered channels consist of continuous metal paths covering the NW. These allow confirmation of the MUX/d-MUX function irrespective of the DUT performance and also provide for reference measurements the MUX and d-MUX series resistance.

Before discussing DUT properties, we analyze the functionality of the MUX/d-MUX circuit. Figure~\ref{Fig3}c shows the conductance of the circuit for each of the 65536 combinations of the first 256 MUX and d-MUX channels. The measurement was performed with positive voltage on the DUT gates, which were therefore all conducting. Indeed, high conductance is observed along the main diagonal, ($\alpha$), which corresponds to both MUX and d-MUX addressing the same DUT channel. This confirms that none of the 1996 SAG NWFETs of the MUX/d-MUX pair fail conduct which would lead to regions of no conductance along the diagonal. In the case of negative $V_\mathrm G$ on the DUT level, every second pixel of the diagonal has $G=0$ (Supplementary Section S5). In the ideal case, the diagonal would be the only non-zero values of the conductance matrix. However, finite off-diagonal conductivity also appears following a repeating pattern every 4, 64, and 128 channels in Fig.~\ref{Fig3}c and d ($\beta, \gamma, \delta$). Since the FETs are conducting at $V_\mathrm G = 0\, \mathrm V$, finite current at these combinations of MUX and d-MUX channels corresponds to rows of NW FETs failing to respond to the gates. This was likely due to a break in gate-lines or a failing bond wire which can occur for large, complex circuits. Figure~\ref{Fig3}e and f schematically illustrates the correlation between the patterns of the matrix and FETs failing to pinch off turn off at various positions in the circuit. For example, a non-responsive gate, on the second d-MUX level from the DUT layer (blue cross in Fig.\ 3e) would allow transport for (MUX,d-MUX) combinations (2,0), (3,1), (6,4), and (7,5) as indicated by blue in Fig.\ 3f. Comparing to the measurement in Fig.~\ref{Fig3}c and Fig.~\ref{Fig3}d the periodically repeating off-diagonal pattern can be assigned to failures of one of the gates in the MUX levels marked with the corresponding labels in Fig.~\ref{Fig3}a. The additional feature appearing at d-MUX channel 192 ($*$) results from the combination of faulty FETs at channels 64 and 128 (Supplementary Section S7 and S8 provides a further analysis of the faults of the circuit). 

Importantly, the MUX/d-MUX configuration allows for identifying and in most cases self-corrects for malfunctioning elements of the SAG circuit; the double redundancy makes the measurement tolerant towards non-symmetrical errors, as, e.g., a non-functioning gate on the MUX side will be intrinsically corrected for by the function of the corresponding d-MUX gate. While errors appearing symmetrically in the MUX and d-MUX side of the circuit cannot be corrected for, they can be identified in the conductance matrix and the corresponding DUT can be excluded from experiments/analysis. This is schematically illustrated in Fig.~\ref{Fig3}e and f: if the MUX/d-MUX FETs fail to pinch-off at the symmetric red/purple positions, addressing levels 0,1,2,3 would also mix signals from levels 4,5,6,7, respectively. Such a situation is readily identified by the symmetric off-diagonal non vanishing elements in the conductance matrix (purple and red in Fig.~\ref{Fig3}f). We note that FETs failing to pinch-off would pass unnoticed in single-ended MUX layouts\cite{smith:2020} where DUT share a common ground. The opposite case, where MUX FETs fail to open would result in periodic non-conductive elements in the diagonal of the matrix. Other examples of MUX/d-MUX circuits are discussed in Supplementary Section S6, showing that even with the amount of failures typical for research-level devices, the self-correcting nature of MUX/d-MUX configuration generally protects against a reduction in the available number of DUT.

As a final comment on MUX operation, we note that bandwidth is a key issue for control electronics. In our experiments, bandwidth was limited by the cryogenic setup, being optimized for low electron temperature, including $\sim 5 \, \mathrm{kHz}$ low-pass filtering of each line. The MUX operation was uninhibited up to these frequencies (Supplementary Section S9), and we expect much higher bandwidth to be possible similar to other InAs NW electronics operating at GHz\cite{tomioka:2012,johansson:2014,egard:2010}.

\section*{Multiplexing of Quantum dot arrays}
Eliminating the device-count road block by the MUX/d-MUX circuit enables fundamentally new experimental approaches in quantum electronics. For example, adding statistical significance to the characterization and optimization of device performance and material properties is crucial for efforts towards up-scaling of quantum circuits. As an example, we demonstrate here the use of the circuit for establishing the statistical reproducibility within large ensembles of lithographically identical devices.  An array of 50 lithographically identical SAG quantum dot devices were embedded in the DUT layer as shown in Fig.~\ref{Fig4}a. Potentials ($V_\mathrm{T}$, $V_\mathrm M$, $V_\mathrm{B}$) applied to three shared gates (top, middle, bottom) simultaneously tune the electrostatics of all 50 devices. This cross-bar approach is an important strategy for limiting the gate-count in up-scaling of QD arrays, however, successful operation requires significant reproducible between devices\cite{li2018crossbar,Borsoi:2022}. Here we benchmark the consistency in the SAG NW QD array by comparing the statistical distributions of QD parameters among devices labeled Dev1-Dev20. First, however, since InAs SAG QDs have thus far not been demonstrated, we establish the characteristics of a single device (Dev1). 

Figure~\ref{Fig4}c shows $G$ vs.\ $V_\mathrm{T}$ and $V_\mathrm{B}$ for fixed $V_\mathrm{M} = 1 \, \mathrm V$. Pinch-off is at $\sim -150 \, \mathrm{mV}$ for both $V_\mathrm T$ and $V_\mathrm B$, and horizontal/vertical structures are attributed to resonances below each gate modulating the transmission\cite{doh:2005}. With both gates near pinch-off, electrons are ideally confined to a NW segment below the middle gate, thus defining a QD. Indeed, fixing $V_\mathrm T$ and $V_\mathrm B$ at the position of the red dot in Fig.\ 4c, diamond shaped regions of low conductance associated with Coulomb blockade (CB) are observed in the map of the differential conductance, $dI/dV_\mathrm{SD}$, vs.\ source bias $V_\mathrm{SD}$ and $V_\mathrm{M}$, as shown in Fig.~\ref{Fig4}b. The $V_\mathrm{SD}$-height of the diamonds provides an estimate of the QD addition energy, being the sum of the electrostatic charging energy $E_\mathrm C$ and the single-particle level spacing $\Delta E$. As discussed in Supplementary Section S10 the QDs have $\Delta E \ll E_\mathrm C$ and from Fig.\ 4b we estimate $E_\mathrm{C} \sim 210 \pm \, 15 \mu \mathrm{eV}$. The capacitance between the QD and the middle plunger gate is estimated as $C_\mathrm{G}=e/\overline{ \Delta V_\mathrm M } = 0.4 \, \mathrm{fF}$ where $ \overline{ \Delta V_\mathrm M } = 0.46\, \mathrm{mV}$ is the average value CB peaks spacings in the range of Fig.~\ref{Fig4}b. This $C_\mathrm{G}$ value agrees with the result ($0.37 \mathrm{fF}$) of simple capacitance estimate based on the gate layout (Supplementary Section S10) and thus support that in this particular gate-configuration, the QD confinement is defined by gates as intended.

To investigate the sensitivity to the tuning, $G(V_\mathrm M)$ was measured at $11 \times 11$ equally spaced $(V_\mathrm T, V_\mathrm B)$ points, spanning the white square in Fig.~\ref{Fig4}c. CB peaks were identified at 108 of the 121 gate-tunings and Fig.~\ref{Fig4}d shows the corresponding map of $\overline{ \Delta V_\mathrm M }$. No systematic trend is observed and the distribution of $\overline{ \Delta V_\mathrm M }$ shown in Fig.~\ref{Fig4}e (top), is symmetric with a mean $\overline{\overline{ \Delta V_\mathrm M }} = 464\, \mathrm{\mu V}$ and standard deviation $\sigma = \pm 27\, \mathrm{\mu V}$, again consistent with a QD defined between the top and bottom gates.

The choice of range for the cross-bar gate-tunings in Fig.\ 4b,d was based on the gate characterization specific for Dev1 (Fig.\ 4c). We now use the MUX circuit to gather statics for the different devices in order to probe the consistency across the array while keeping these same tuning parameters. Values of $\overline{ \Delta V_\mathrm M }$ were extracted from $G(V_\mathrm M)$ traces measured for all devices and all gate-points. All devices were operational and exhibited coulomb blockade and from the resulting 2420 gate-traces, 17924 CB peaks were identified and fitted. Examples of measured data and peak analysis are presented in Supplementary Section S12 and S13. The distributions of $\overline{\Delta V_\mathrm{M}}$ for all devices are included in Supplementary Section S14 and examples for Dev2-5 are shown in Fig.\ 4e. A comparison of the distributions and their mean values, $\overline{\overline{ \Delta V_\mathrm M }}$, among the devices of the array, is shown in Fig.\ 4f. Except for Dev7, the distribution means fall within one $\sigma$ of the overall common mean. The spread between devices could be affected by structural variations between nanowires due to SAG processing or to variation in post growth device processing. 
The spread within each devices could be related to changes in the effective confinement potential with gate tunings and may be different between devices due to random impurity in the vicinity of the devices. 

Finally, Fig.~\ref{Fig4}g shows the joint distribution of all $\overline{ \Delta V_\mathrm M }$ and Fig.~\ref{Fig4}h illustrates the number of devices displaying CB for all measured combinations of $V_\mathrm{T}$, $V_\mathrm{B}$. In 27 out of the 121 point in cross-bar gate space, \textit{all} 20 devices simultaneously exhibiting CB. The circles mark the 7 tunings where \textit{all} devices fulfill the stricter criterion of showing CB and having $\overline{ \Delta V_\mathrm M }$ within $\pm 2\sigma$ of the joint mean peak spacing. Figure 4f-h constitute a key result of the current study, establishing both a level of device to device reproducibility supporting the potential of SAG for as a scalable platform for quantum electronics. Further, the statistical bench-marking of the QD devices explicitly demonstrates a key example of the new possibilities enabled by the integration of MUX/d-MUX circuits.

\section*{Conclusions and Outlook}

In conclusion, we successfully fabricated and operated cryogenic multiplexers/de-muliplexer circuits based on InAs NWs grown bottom-up by selective area growth. The circuit removes the limitations on device count in conventional cryogenic electronics thus enabling new experimental strategies such as searches through large ensembles of devices for rare or exotic phenomena, establishing the correlation between device performance and materials properties or device geometry, and allows establishing the statistical reproducibility among devices - a prerequisite for further scaling quantum of circuits. This capacity was demonstrated by statistically characterizing an ensemble of SAG quantum dots. In general, the methods developed here enable optimization of quantum materials and devices based on automated acquisition of statistically significant datasets rather than proof-of-principle examples. This direction will be empowered by the ongoing developments of advanced data evaluation\cite{lesage:2015} and machine learning\cite{zwolak:2020, moon:2020, nguyen:2021} for unsupervised and optimized acquisition and tuning of large ensembles of quantum devices with many tuning parameters\cite{lennon:2019,darulova:2020,chatterjee:2022}. The circuit may be expanded further by replacing the single lines in the current design by a multi-channel bus\cite{ward:2013} to enable, e.g., integration of charge sensors, multi-terminal devices, complex gate-architectures, and/or the operating the MUX as a multi-channel DAC\cite{puddy:2015}.

\section*{Methods}
For SAG fabrication and synthesis, a $10 \, \mathrm{nm}$ \ce{SiO2} mask layer was first deposited on epi-ready GaAs \hkl(311)A substrates by plasma enhanced chemical vapour deposition. $0.15 \times 10 \, \mu \mathrm m$ rectangular openings were defined in the oxide along the \hkl[0-11] direction by e-beam lithography (EBL) and dry etching. The openings were arranged in $512 \times 16$ arrays with a pitch of $2 \, \mu \mathrm m$ and $20 \, \mu \mathrm m$ along the \hkl[0-11] and \hkl[-233], respectively (Fig.~\ref{Fig1}d). GaAs(Sb)/InAs double layer NWs were selectively grown in the openings where the GaAs(Sb) buffer was introduced to improve the crystal surface for the subsequent InAs transport channel\cite{Krizek2018Sep}. Synthesis details and structural analysis are provided in Supplementary Sections S1 and S2. For device fabrication, Ti/Au ohmic contacts to the SAG NWs were defined on the growth substrate by standard EBL, metal evaporation, and liftoff. Subsequently, a $15 \, \mathrm{nm}$ \ce{HfO2} gate dielectric was deposited by atomic layer deposition and top-gates were defined by electron beam lithography, metal evaporation, and liftoff. The QD devices in Fig.\ 4 have a contact separation of 900 nm, top/middle/bottom cross-bar gates are 300/200/300 nm wide and have a 150 nm spacing. Electrical measurements were carried out in a dilution refrigerator with a base temperature of $20 \, \mathrm{mK}$. The conductance, $G=I/V_\mathrm{SD}$, where $I$ is the drain current generated by source voltage $V_\mathrm{SD}$, was measured as a function of the gate potential, $V_\mathrm G$, using standard lock-in techniques. The series resistance $R_\mathrm S$ for data presented in Fig.~\ref{Fig2} was estimated by fitting the $G(V_\mathrm G)$ traces with the standard expression\cite{gul2015towards} $G = \left( R_\mathrm S + \frac{L^2}{\mu C (V_\mathrm G - V_\mathrm{TH})} \right)^{-1}$ where $L$ is the gate length, $C$ is gate capacitance simulated as described in Supplementary Section S3 and S4, and $\mu_\mathrm{FE}$ is the electron mobility. When sweeping $V_\mathrm M$ in the QD measurements, the barrier gates were compensated for a slight capacitive cross coupling (Supplementary Section S10).

\subsection*{Data availability} Full data sets for all figures, STEM micrographs, transport data, electronic logbooks and other data that support the findings of this study are available  online at 
\begin{verbatim}
https://sid.erda.dk/
\end{verbatim}

\subsection*{Acknowledgements}
This research was supported by research grants from Villum Fonden (00013157) and the European Research Council (866158). ICN2 acknowledges funding from Generalitat de Catalunya 2021SGR00457. The authors thank support from “ERDF A way of making Europe”, by the “European Union”. ICN2 is supported by the Severo Ochoa program from Spanish MCIN / AEI (Grant No.: CEX2021-001214-S) and is funded by the CERCA Programme / Generalitat de Catalunya. Authors acknowledge the use of instrumentation as well as the technical advice provided by the National Facility ELECMI ICTS, node "Laboratorio de Microscopías Avanzadas" at University of Zaragoza. We acknowledge support from CSIC Interdisciplinary Thematic Platform (PTI+) on Quantum Technologies (PTI-QTEP+).

\subsection*{Competing interests}
The authors declare no competing financial interests

\subsection*{Supporting Information} Supporting information contains extended information on sample preparation and MBE growth, NW structure investigation via GPA, extended transport datasets for Fig.~\ref{Fig4}b, details of the fitting procedures employed for Fig.~\ref{Fig4}c and d, as well as a more in-depth description of the MUX/d-MUX circuit benchmarking, measurement bandwidth, and switching speed. 

\raggedright
\bibliographystyle{naturemag}
\bibliography{ref,TSJZotero}

\end{document}


\author{Dags Olsteins}
\affiliation{Center For Quantum Devices, Niels Bohr Institute, University of Copenhagen,\\ 2100 Copenhagen, Denmark}

\author{Gunjan Nagda}
\affiliation{Center For Quantum Devices, Niels Bohr Institute, University of Copenhagen,\\ 2100 Copenhagen, Denmark}

\author{Damon J. Carrad}
\affiliation{Department of Energy Conversion and Storage, Technical University of Denmark, 2800 Kgs.Lyngby, Denmark}

\author{Daria V. Beznasiuk}
\affiliation{Department of Energy Conversion and Storage, Technical University of Denmark, 2800 Kgs.Lyngby, Denmark}

\author{Christian E. N. Petersen}
\affiliation{Department of Energy Conversion and Storage, Technical University of Denmark, 2800 Kgs.Lyngby, Denmark}

\author{Sara Martí-Sánchez}
\affiliation{Catalan Institute of Nanoscience and Nanotechnology (ICN2), CSIC and BIST, Campus UAB, Bellaterra, Barcelona, Catalonia, Spain}

\author{Jordi Arbiol}
\affiliation{Catalan Institute of Nanoscience and Nanotechnology (ICN2), CSIC and BIST, Campus UAB, Bellaterra, Barcelona, Catalonia, Spain}
\affiliation{ICREA, Passeig de Lluís Companys 23, 08010 Barcelona, Catalonia, Spain}

\author{Thomas Sand Jespersen}
\affiliation{Center For Quantum Devices, Niels Bohr Institute, University of Copenhagen,\\ 2100 Copenhagen, Denmark}
\affiliation{Department of Energy Conversion and Storage, Technical University of Denmark, 2800 Kgs.Lyngby, Denmark}
\email{tsaje@dtu.dk}

\title{Supplementary Information \\ Cryogenic Multiplexing with Bottom-up Nanowires}

\maketitle 

\section{Details of selective area growth}
The details of the selective area growth (SAG) using MBE are provided here. After preparing the substrate for SAG, the sample is introduced to the MBE system. The sample was degassed in the loadlock for 4 hours at $200\degree$\,C and then transferred to a second chamber connecting the loadlock to the main chamber. Here, the sample was further degassed at a heating station for 1 hour at $400\degree$\,C. After transfer to the MBE growth chamber, the substrate was thermally annealed by increasing the substrate temperature to $T_\text{sub} = 400\degree$\,C with a ramp rate of $20\degree$\,C/min and then further to $T_\text{sub} = 610\degree$\,C with a ramp rate of $10\degree$\,C/min under a constant As$_4$ beam equivalent over-pressure of $1.4 \times 10^{-5}$\,mbar. RHEED was used to monitor the removal of the native oxide at this temperature for approximately 13 minutes (this time varies slightly from sample to sample), the last minute of which the intensity of the specular RHEED spot does not increase\cite{Isu1998Jun}. Subsequently, $T_\text{sub}$ was reduced to the GaAs(Sb) buffer growth temperature of $T_{\text{sub},\text{GaAs(Sb)}} = 600\degree$\,C. For the buffer growth, a As/Ga ratio of 9 and Sb/Ga ratio of 3 was used. This ensures that the growth rate is dependent on the Ga flux and the growth is performed with a Ga growth rate of $0.1$\,ML/s for $30$\,min. The substrate temperature was further reduced to $T_{\text{sub},\text{InAs}} = 500\degree$\,C and stabilized during $10$\,min for InAs growth. An As/In ratio of 9 and an In growth rate of $0.06$\,ML/s was used and InAs was grown for $18$\,min. The choice of a lower substrate temperature and a slower growth rate is employed to minimize temperature assisted Ga diffusion into the InAs channel\cite{Beznasyuk2022Mar}. The substrate temperature was monitored before, shortly at the beginning and after the growth using a pyrometer and the variation in $T_\text{sub}$ was maximally $2\degree$\,C. This uncertainty is attributed to the variation in temperature read-out caused by radiation from different regions of the substrate. 

\section{Crystal Structure, Strain and Composition}

\begin{figure}[htb]
  \centering
    \includegraphics[width = \linewidth]{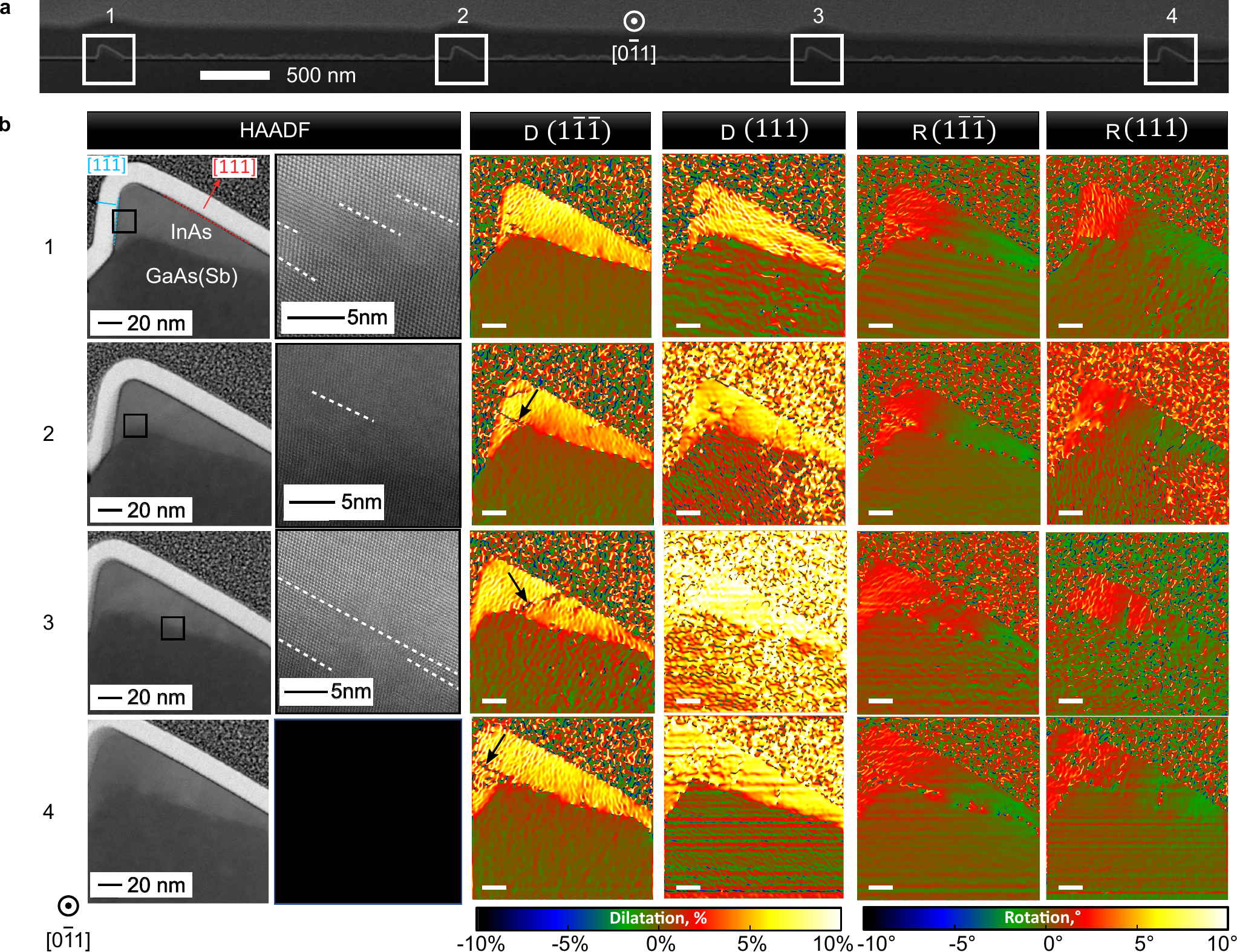}
    \caption{\textbf{a} A cross-sectional view of 4 InAs/GaAs(Sb) NWs roughly from the middle of an array of 50 NWs oriented along the \hkl[0-11] direction on a GaAs\hkl(311)A substrate. \textbf{b} (column 1) High-angle annular dark field scanning tunneling micrographs (HAADF-STEM) show the GaAs(Sb) and InAs regions of the NWs. The morphology of all 4 NWs is uniform and the InAs channel exhibits \hkl{111}A facets. \textbf{b} (column 2) Stacking faults originating at the GaAs(Sb)/InAs interface are highlighted with white dashed lines. \textbf{b} (column 3, 4) Dilatation maps of all NWs show an array of dislocations at the GaAs(Sb)/InAs interface, and arrows point to stacking faults. \textbf{b} (column 5, 6) The rotational maps indicate rotation of the crystal planes between the GaAs(Sb) buffer and the InAs channel. Scale bars correspond to $20$\,nm in all dilatation and rotation maps.
    } 
    \label{Supp_HAADF}
  \hfill
\end{figure}

Figure~\ref{Supp_HAADF} shows the morphology, strain and crystal structure of the NWs comparable to NWs investigated in this study. A uniform morphology is visible for 4 InAs/GaAs(Sb) NWs oriented along the \hkl[0-11] direction on a GaAs\hkl(311)A substrate in Fig~\ref{Supp_HAADF}a. The InAs channel exhibits \hkl{111}A facets. It should be noted that the cross-sectional shape is specific to the choice of substrate, in-plane orientation, mask dimensions such as the width and pitch, as well as the growth parameters listed in the previous section. 
 
Figure~\ref{Supp_HAADF}b shows the crystal structure of all 4 NWs, along with the corresponding geometric phase analysis (GPA) of the dilatation and rotation maps. Figure~\ref{Supp_HAADF}b (column 2) shows stacking faults (highlighted as white dashed lines) originating at the GaAs(Sb)/InAs interface and propagating towards the InAs channel. It can be observed the dilatation maps in Fig.~\ref{Supp_HAADF}b (column 3, 4) that the main relaxation occurs through creation of an array of dislocations at the GaAs(Sb)/InAs interface. The rotational maps in Fig.~\ref{Supp_HAADF}b (column 5, 6) show rotation in the crystal planes which comes along as an elastic relaxation mechanism.

\begin{figure}[htb]
  \centering
    \includegraphics[width = 0.5 \linewidth]{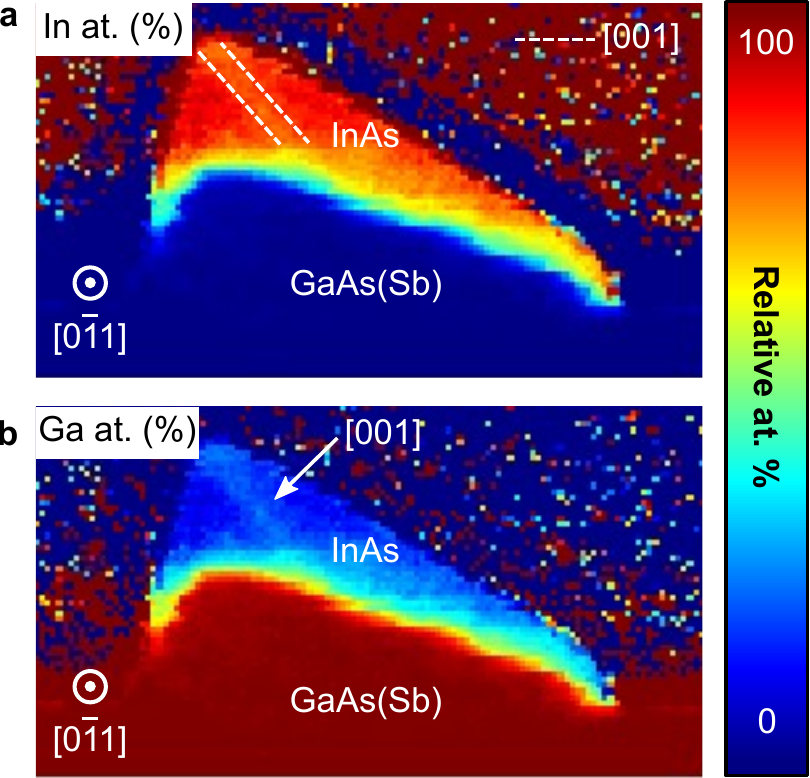}
    \caption{\textbf{a} In and \textbf{b} Ga relative atomic \% compositional map of a typical NW used in this study obtained by electron energy loss spectroscopy (EELS). The InAs region is found to be diluted with $8-10\%$ Ga. The white dashed lines in \textbf{a} show the Ga-rich region within the InAs channel along which Ga is found to diffuse more than in the surrounding region. The preferential diffusion of Ga along the \hkl[001] direction is better visible in \textbf{b} where the map is relative to the Ga atomic \%.
    } 
    \label{Supp_HAADF-2}
  \hfill
\end{figure}

Electron Energy Loss Spectroscopy (EELS) of the NWs is used to obtain the elemental composition of the NW. In particular, relative elemental quantification of the In vs Ga ratio in atomic \% is shown in  Fig.~\ref{Supp_HAADF-2}a and b respectively. Ga diffusion towards the InAs channel of ~8-10 \% is observed along the \hkl[001] direction. The diffusion region is highlighted in Fig.~\ref{Supp_HAADF-2}a between the white dashed lines and is better visible in Fig.~\ref{Supp_HAADF-2}b. Such preferential diffusion has been reported in GaAs(Sb)/InGaAs/InAs NWs grown on GaAs\hkl(001) substrates \cite{Beznasyuk2022Mar}. It has been proposed that Ga diffusion arises as a thermally activated strain minimization mechanism during growth of lattice-mismatched InAs and GaAs. It should be noted that an In rich region only exists at the outermost layers of the conduction channel whereas the region in between is diluted with the diffused Ga. This can be circumvented by further reducing the substrate temperature during InAs growth, but is not attempted because it comes at the cost of abating the selectivity of the sample.

\clearpage
\section{Electrical characterization of SAG Nanowire FETs.}

\begin{figure}[htb]
  \centering
    \includegraphics[width = 0.9\linewidth]{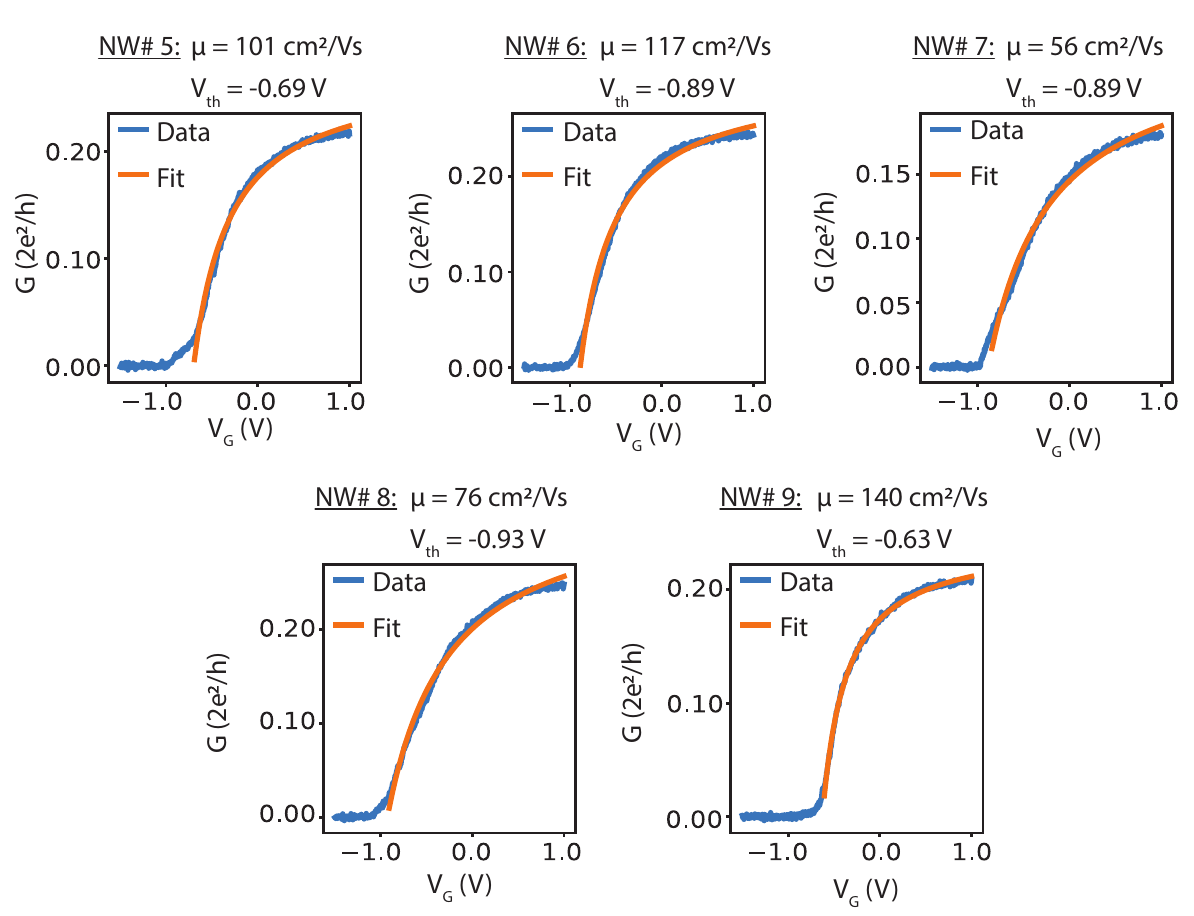}
    \caption{Conductance vs.\ gate potential for 5 SAG nanowire FETs. The devices are similar to Fig.\ 2 of the main manuscript and are fabricated on 5 consecutive single SAG NWs of the arrays. The measurements were performed at 100K and the parameters extracted from the fit are included above in each case.
    } 
    \label{gulfit}
  \hfill
\end{figure}

Figure \ref{gulfit} shows measurements of the conductance vs.\ gate potential for 5 NW FETs similar to those discussed in Fig.\ 2 of the main manuscript. Here, only a single NW is measured $(M=1)$ and 5 consecutive wires of the SAG array were chosen. The channel length was $1 \, \mu \mathrm m$ and measurements were performed at $100 \, \mathrm K$. In each case the conductance, $G(V_\mathrm G)$, was fitted to the model of Ref.\ \cite{gul:2015}, $G(V_\mathrm G)^{-1} = R_\mathrm S + L^2/(\mu C(V_\mathrm{G} - V_\mathrm{TH} ))$ where $R_\mathrm S, \mu, V_\mathrm{TH}$ and $C$ are the contact/series resistance, mobility, threshold voltage and gate-NW capacitance respectively. We take $C= 5.3 \mathrm{fF}$ corresponding to the value extracted from the simulation described in Section \ref{SECT:model}. The resulting mobility and threshold voltage for each device is stated in the figure. The mobility is between 56 and 140 $\mathrm{cm^2/Vs}$ and $V_\mathrm{TH}$ is between -0.63\, V and -0.93\, V consistent with the discussion in the main text of un-gated FETs not being in pinch-off. The values are relatively low for InAs which we attribute to the stacking faults and dislocations associated with the unbuffered growth of InAs on non-lattice matched GaAs\cite{beznasyuk:2022}. 

\clearpage
\section{Modelling Gate Capacitance of NWFETs} \label{SECT:model}

\begin{figure}[htb]
  \centering
    \includegraphics[width = \linewidth]{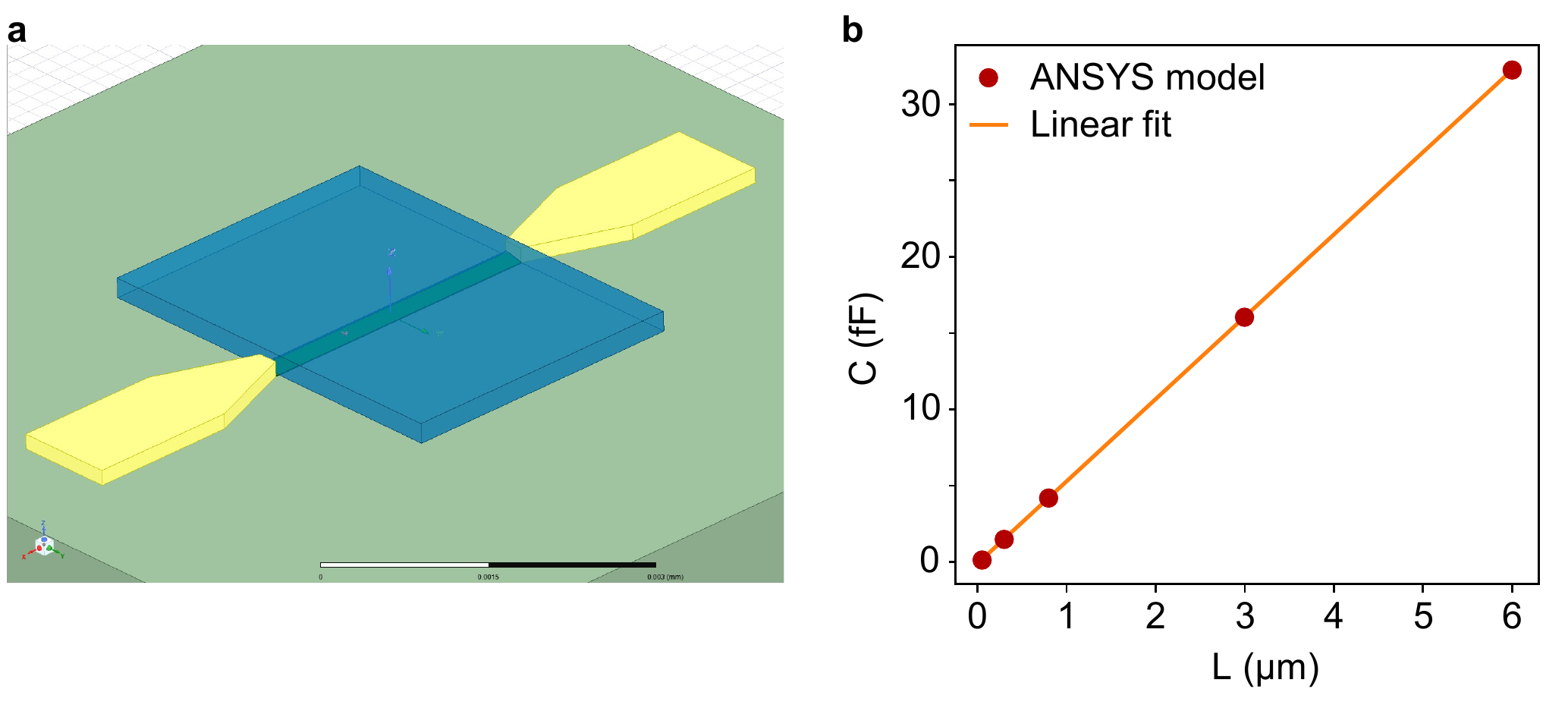}
    \caption{\textbf{a} 3D model of a NWFET device with a $3\, \mathrm{\mu m}$ long \ce{InAs} channel rendered in ANSYS software based on AFM, SEM, and TEM measurements. Ti/Au contacts in yellow, Ti/Au topgate in blue \textbf{b} Capacitance values calculated by the model for different \ce{InAs} channel lengths and linear fit to the results.
    } 
    \label{FIG:ansys}
  \hfill
\end{figure}

Gate capacitance for NWFET devices is estimated using the ANSYS\textsuperscript{\textregistered{}} Electronics Desktop 2021 R2 software. Model NWFET devices are rendered based on measurements from AFM, SEM, and cross-section TEM for lengths $50\, \mathrm{nm}$, $300\, \mathrm{nm}$, $800\, \mathrm{nm}$, $3\, \mathrm{\mu m}$, and $6\, \mathrm{\mu m}$ of the semiconductor NW segment. An example model is shown in Fig. \ref{FIG:ansys}a. The Maxwell3D module of the software models electrostatics using finite element analysis and solves Maxwell's equations in a finite region of space. From the resulting values shown in Fig.\ \ref{FIG:ansys}b a capacitance per unit length of $\sim 5.3\, \mathrm{fF / \mu m}$ is extracted to find the capacitance for other NW segment lengths.

\clearpage
\section{Diagonal of the conductance matrix with $V_\mathrm G = -1 \, \mathrm V$}

\begin{figure}[htb]
  \centering
    \includegraphics[width = 0.9\linewidth]{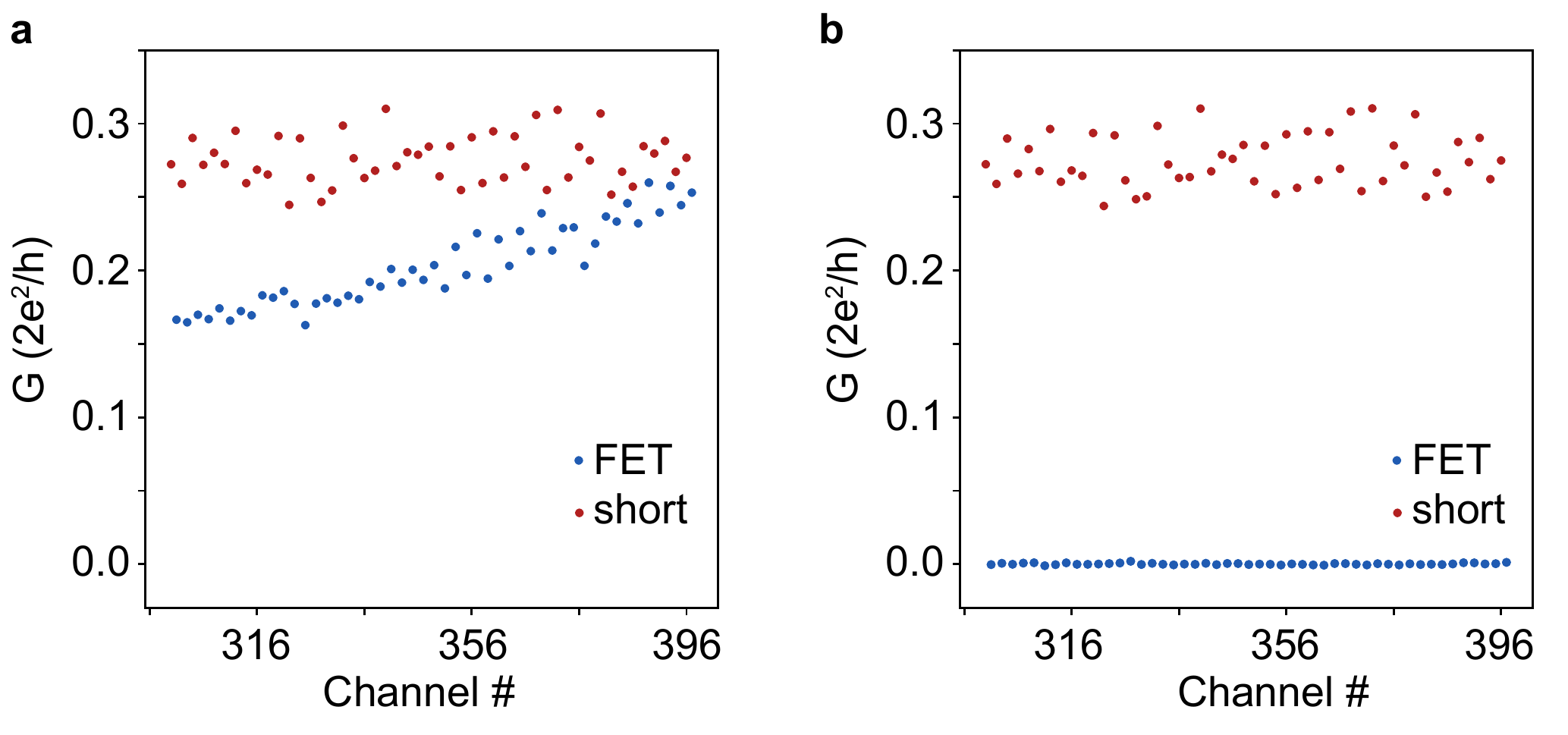}
    \caption{\textbf{a (b)} Conductance of the MUX/d-MUX circuit when both MUX and d-MUX addresses the same channels \#300 to \#398 with $V_\mathrm{G} = +1\, \mathrm{V}$ ($V_\mathrm{G} = -1\, \mathrm{V}$). Odd channels addresses NW FETS at the DUT level while even channels are shorted. 
    } 
    \label{evenodd}
  \hfill
\end{figure}

Figure \ref{evenodd}a,b shows the conductance along the diagonal of the conductance matrix of Fig.\ 3c for MUX/d-MUX channels \#300 to \#398 for $V_\mathrm G = 1 \, \, \mathrm V$ and $V_\mathrm G = -1\, \mathrm V$, respectively. Odd channels addresses NW SAG FETS at the DUT level while even channels are shorted. The shorted channels are unaffected by $V_\mathrm G$ while the FETs have $G=0 \, \mathrm{2e^2/h}$ at negative $V_\mathrm G$. For the conductance matrix in Fig.\ 3c of the main manuscript, this corresponds to every second pixel along the diagonal showing no conductance as discussed in the main manuscript. 

\clearpage
\section{Additional Multiplexer Circuits}

\begin{figure}[htb]
  \centering
    \includegraphics[width = 0.9\linewidth]{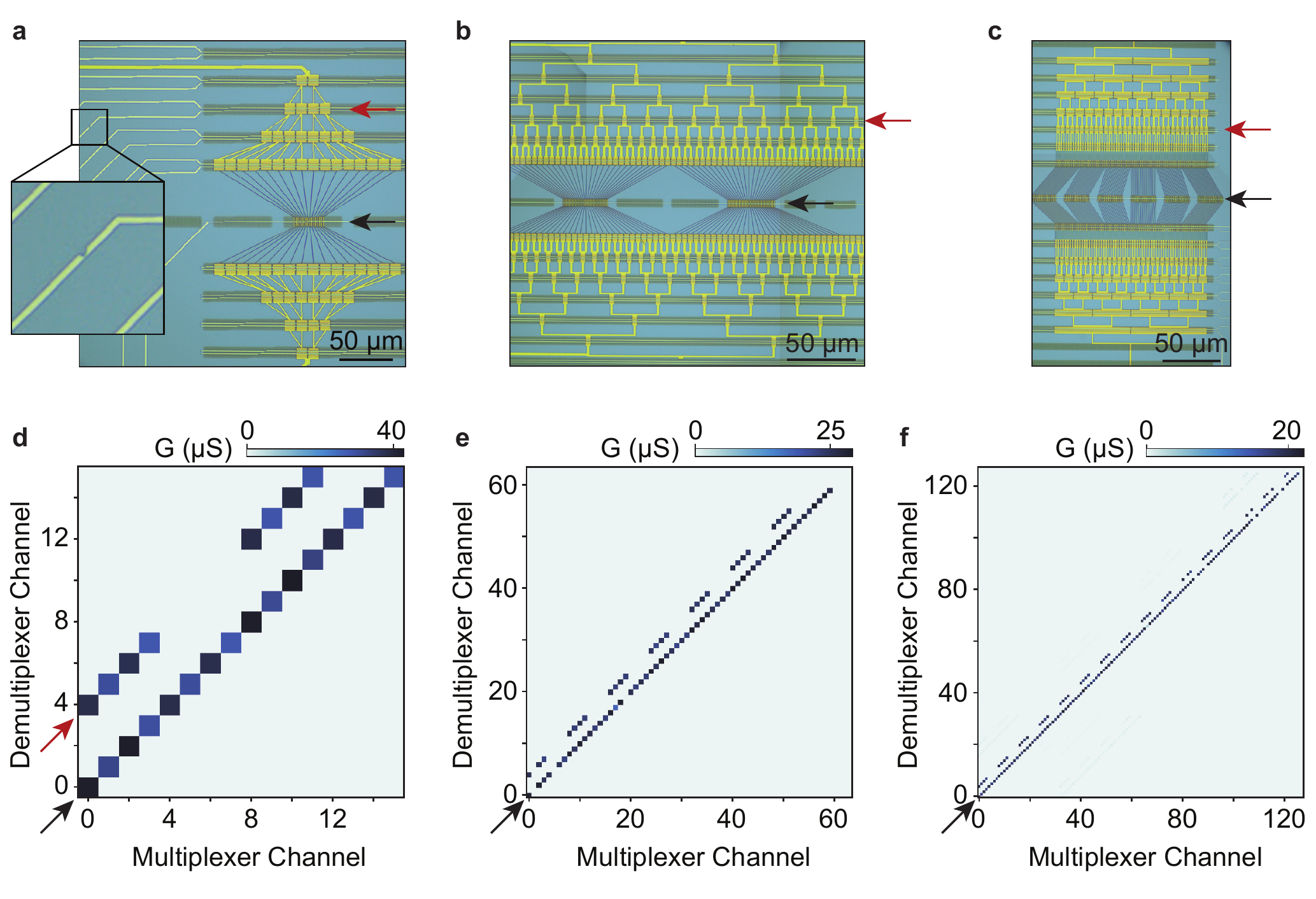}
    \caption{\textbf{a-c} stitched optical microscope micrographs of three MUX/d-MUX circuits with 16, 64, and 128 outputs respectively. Red arrows indicate the level responsible for the off diagonal features in d-f. \textbf{d-f} the corresponding conductance matrices.
    }
    \label{Supp_oldmpx}
  \hfill
\end{figure}

Other multiplexer circuits were created in the scope of the research presented here. Fig.\ref{Supp_oldmpx} shows stitched optical microscope micrographs and the correpsonding conductance matrices of three additional MUX/d-MUX circuits fabricated on another chip from the same nanowire growth as the one shown in the main text. All three designs are fully functional and have 16, 64, and 128 outputs respectively. A broken gate line (inset in Fig. \ref{Supp_oldmpx}a) is responsible for the off-diagonal features in the conductance matrices in agreement with simulations described in S7.

\clearpage
\section{Conductance Matrix Simulations}

\begin{figure}[h!]
  \centering
    \includegraphics[width = 0.8\linewidth]{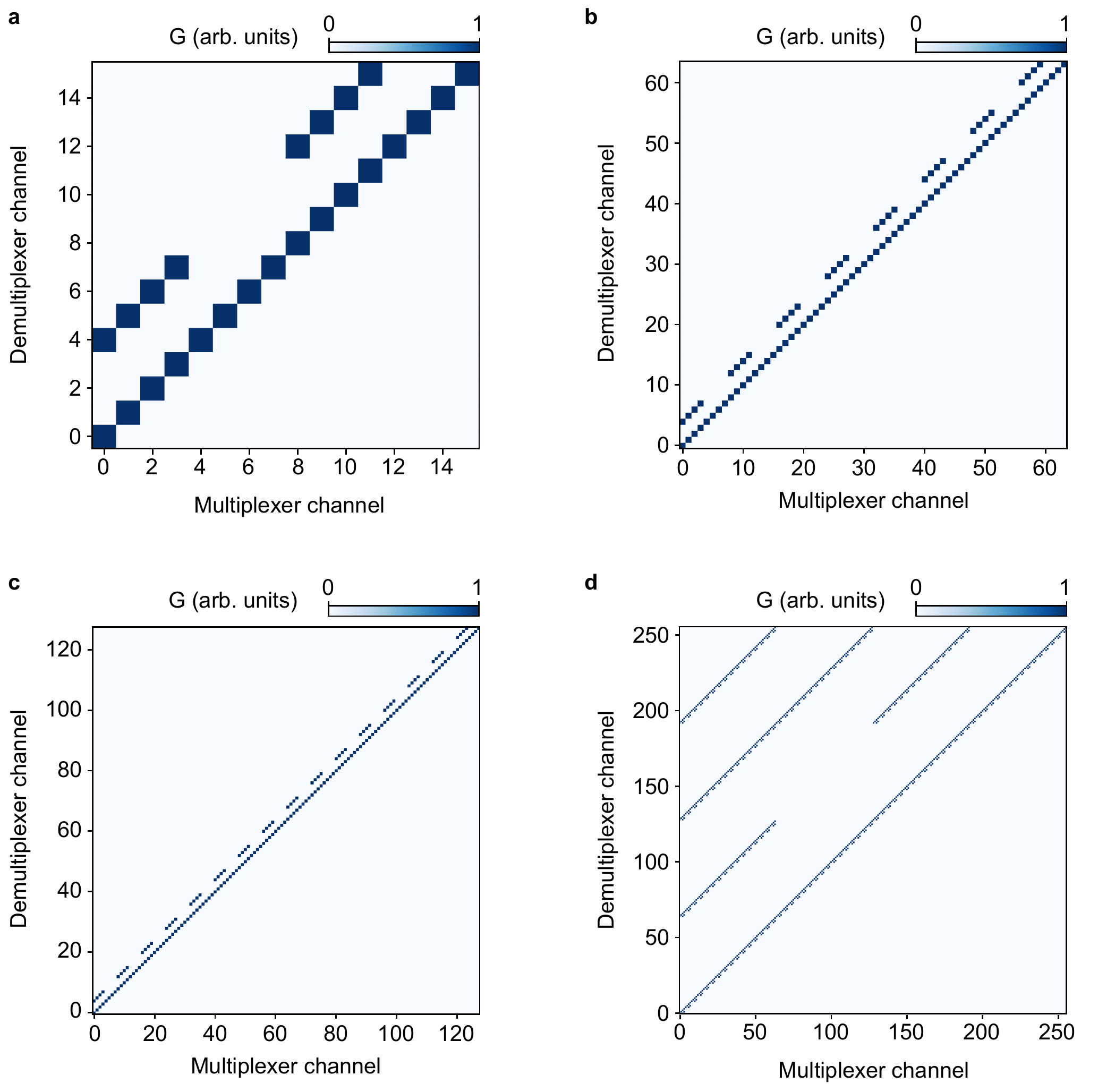}
    \caption{\textbf{a-c} Simulated conductance matrices for the circuits shown in Fig. \ref{Supp_oldmpx}. The gate identified in the inset in Fig. \ref{Supp_oldmpx} is assumed to be broken. \textbf{d} Simulated conductance matrix for the circuit in main text Fig. 3. Gate failures chosen to match the observed matrix pattern.
    } 
    \label{Supp_sims}
  \hfill
\end{figure}

To verify that the patterns observed in the conductance matrices agree with the assumed gate failures a Monte Carlo simulation was constructed. The simulation takes the number of levels in the multiplexer $L$ as an input and gate failures on both the MUX and d-MUX side can be specified. The simulation assumes that a failing gate is non-functional across the entire level. The algorithm then simulates a random walk from source to drain of $2^L \times 10$ walkers for all MUX and d-MUX channel combinations. If a walker encounters a FET in depletion it is discarded, if at least one walker reaches the end of the circuit, a value of 1 is assigned to the corresponding matrix element. Figure \ref{Supp_sims} shows simulated conductance matrices of the circuits shown in Fig. \ref{Supp_oldmpx} (Fig. \ref{Supp_sims} a-c) and main text Fig.\ 3 (Fig. \ref{Supp_sims} d). For the simulations in Fig.\ \ref{Supp_sims}a-c it was specified that the gate highlighted in the inset of Fig. \ref{Supp_oldmpx}a is broken and for the simulation in Fig.\ \ref{Supp_sims}d it was specified that gates are broken in levels 0 and 6 in the MUX side and on level 7 in the d-MUX side. However, it must be noted that identical matrices can result from different combinations of broken gates.

\clearpage
\section{Examples of conductance matrices of Multiplexers containing multiple faulty FET gates}

\begin{figure}[htb]
  \centering
    \includegraphics[width = 0.9\linewidth]{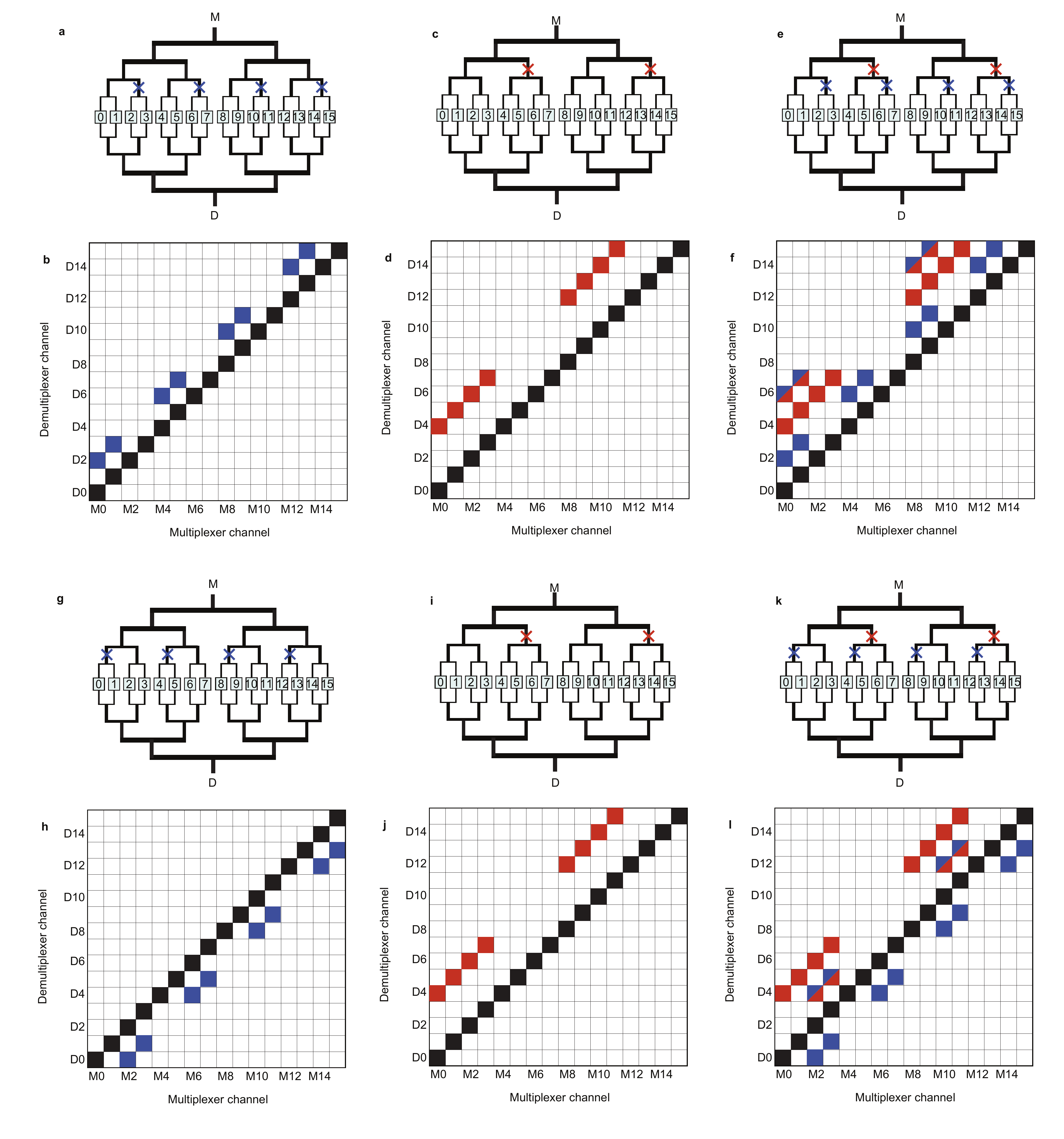}
    \caption{Examples of conductance matrices for a 16x16 multiplexer/demultiplexer circuit containing FET rows failing to pinch-off at various locations in the circuit. The examples illustrates the non-obvious consequences of multiple faults combined (panels e,f and k,l).} 
    \label{MPXcompoundErrors}
  \hfill
\end{figure}

In Fig.\ 3e,f of the main manuscript, the consequences for the conductance matrix of individual FETs or FET rows failing to pinch-off at various locations of the MUX/d-MUX circuit were discussed. When multiple FETs or FET rows fail at the same time, their consequences for the conductance matrix do not just add but are coupled. As an example, the purple arrow in Fig.\ 3c of the main manuscript, appears for multiplexer channel 0 and demultiplexer 192. Since 192 is not a power of 2, such a feature cannot be related to the failing of a FET row. Rather, 192=128+64 and is a consequence of the combined faults located at row 6 $(64=2^6)$ and 7 $(128=2^7)$ (the FETs closest to the DUT layers is nr.\ 0). To illustrate such coupling explicitly, Fig.\ \ref{MPXcompoundErrors}a-e shows the consequences of errors at layer 1 (periodicity 2) and 2 (periodicity 4) and their combination (periodicity 2,4 and 6). Here, the colored crosses on the schematic circuit indicate a FET failing to pinch-off (i.e.\ always open), and the resulting non-vanishing elements of the conductance matrix are shown below. In these schematics, the DUT layer is considered fully conducting. If faults occur on opposing branches or on opposing sides of the multiplexer circuit, the compound errors result in a more intricate pattern as shown in Fig.\ \ref{MPXcompoundErrors}g-l.

\clearpage
\section{Measurement bandwidth}
\begin{figure}[htb]
  \centering
    \includegraphics[width = 0.85\linewidth]{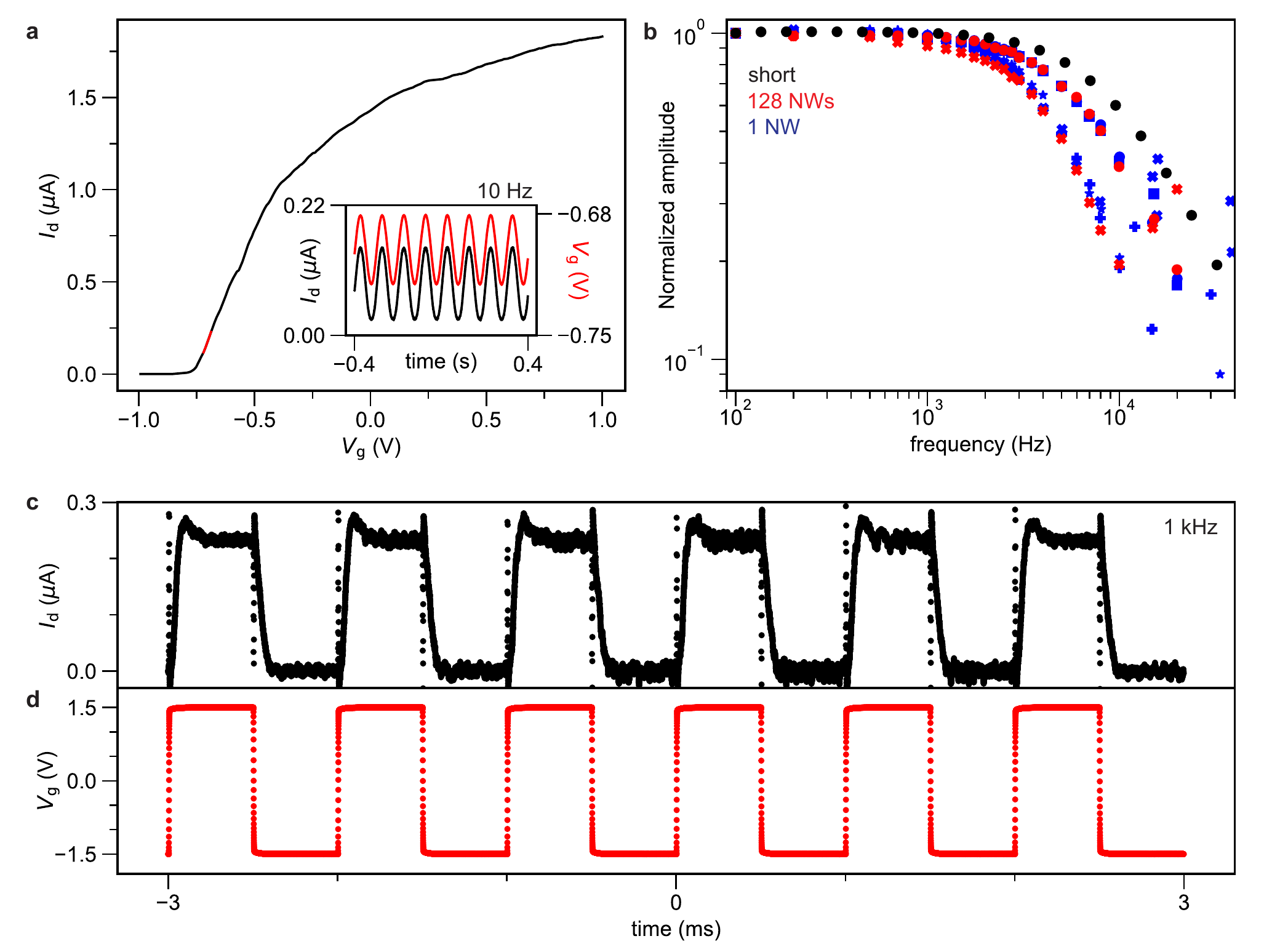}
    \caption{(a) Drain current $I_\mathrm{d}$ at fixed dc source-drain voltage $V_\mathrm{sd}=0.1$~V in response to gate voltage $V_\mathrm{g}$ with no ac component. Inset: $I_\mathrm{d}$ modulation vs time (black) in response to $V_\mathrm{g}=-0.7+0.04$sin$(20\pi t)$~V (red), i.e., an ac voltage at 10~Hz with amplitude $V_\mathrm{ac}=0.04$~V, offset by $V_\mathrm{dc}=-0.7$~V to span the red section of the trace in the main panel. (b) Amplitude of the ac-component of $I_\mathrm{d}$ normalised to the amplitude at 100~Hz as a function of the frequency for two shorted cryostat lines (black), two transistors with 128 SAG NWs (red), and five transistors consisting of a single SAG NW (blue). The cut-off frequency is set by the cryostat wiring. (c) $I_\mathrm{d}$ response of a single SAG NW to a (d) 3~V, 1~kHz square wave applied to the gate.} 
    \label{SAGBandwidth}
  \hfill
\end{figure}

All measurements were conducted in a dilution cryostat optimised for low electron temperature by including 5-stage RC and 21-stage pi-filters on each measurement line. The rejection of high-frequency radiation assists in obtaining a low electron temperature but naturally limits the measurable bandwidth to a few kHz. While this prevented us from benchmarking our InAs transistors at frequencies up to $\sim 100$ GHz as previously demonstrated,\cite{tomioka:2012,egard:2010,johansson:2014} we show in Fig.~\ref{SAGBandwidth} that MPX elements consisting of both one and 128 SAG NWs were capable of operating at frequencies up to the cut-off set by the cryostat filtering. To do this, a sinusoidal signal, $V_\mathrm{ac}$sin$(\omega t)$ was applied to the gate of the selected MPX element, offset with a dc voltage $V_\mathrm{dc}$ such that the transistor was operating in a linear region of drain current $I_\mathrm{d}$ vs $V_\mathrm{dc}$ (Fig.~\ref{SAGBandwidth}a). This ensured high-fidelity transduction (Fig.~\ref{SAGBandwidth}a, inset). Fig.~\ref{SAGBandwidth}b shows the amplitude of the resulting sinusoidal $I_\mathrm{d}$ component plotted as a function of frequency $f=\omega/2\pi$, normalised to the amplitude at 100~Hz. Shown also is the frequency response of two cryostat lines shorted on another sample board loaded during the same cooldown. The frequency responses fall along three paths, correlated with the three different break-out boxes and looms that the lines were grouped in. This indicates that the observed 3-8~kHz low-pass cut-off is set by the cryostat filtering and wiring, with the multiplexer operating without degradation until this point.

To demonstrate the multiplexer switching capabilities within these limits, Fig.~\ref{SAGBandwidth}c shows the MPX drain current as a function of time in response to a 1~kHz, 3~V square wave applied to the gate of a single transistor in the MPX, centered around $V_\mathrm{g}=0$ (Fig.~\ref{SAGBandwidth}d). The transistor switches from fully open to fully closed, with the output stable after the filter-defined $\sim 0.1$~ms rise time. While this represents the upper limit of operation speed in our current set-up, future experiments are planned to measure the bandwidth of SAG devices without these limitations.

\clearpage 
\section{QD Capacitances and Gate Cross-coupling}

\begin{figure}[htb]
  \centering
    \includegraphics[width = 14 cm]{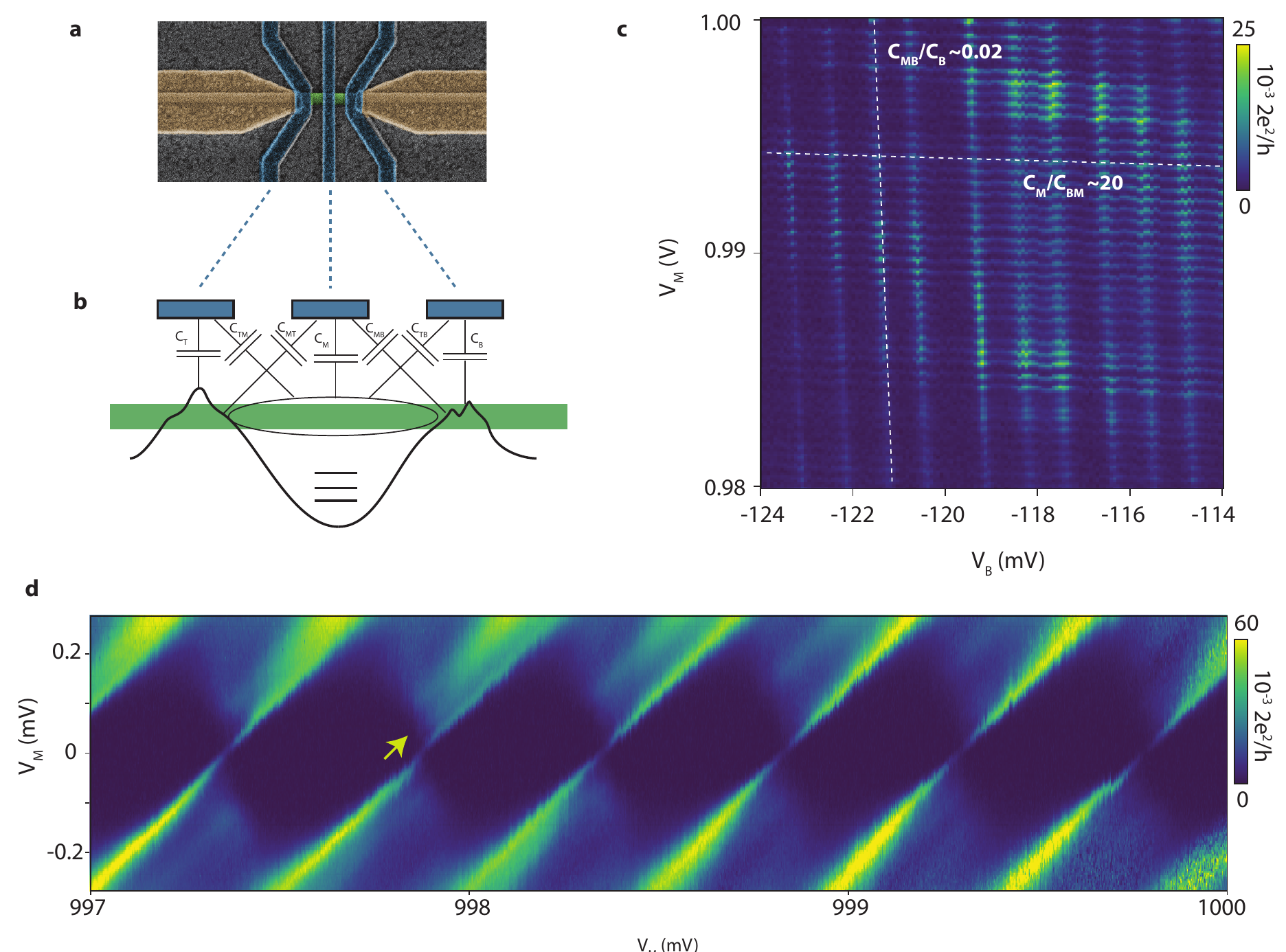}
    \caption{\textbf{a} SEM of a single QD device of the array. \textbf{b} Illustratino of the gate cross-couplings. \textbf{c} Measurement device conductance vs.\ $V_\mathrm M$ and $V_\mathrm B$. The fine structure which is nearly independent of $V_\mathrm V$ corresponds to the CB peaks of the QD in the middle segment, while vertical structures corresponds to resonances below the barrier gate. The slopes yields the ratio of gate-couplings as indicated. \textbf{d} Typical bias spectrum with indication of excited state.} 
    \label{Supp_coupl}
  \hfill
\end{figure}

Figure \ref{Supp_coupl}\textbf{a} shows a zoom of a single NW with the three gates used for tuning QD in the main text and panel \textbf{b} schematically shows the couplings taken into account. The capacitance $0.37\, \mathrm{fF}$ from the middle gate to the NW QD was estimated by a simple parallel plate capacitor taking typical values $\varepsilon_{HfO}=18$, oxide thickness of $15\, \mathrm{nm}$, nanowire width of $180\, \mathrm{nm}$, and a gate width of $200\, \mathrm{nm}$. As seen in Fig.\ 4\textbf{c} of the main text there is negligible coupling between the gates $V_\mathrm{T}$ and $V_\mathrm{B}$. Figure \ref{Supp_coupl}c shows a typical measurement of the conductance vs.\ $V_\mathrm M$ and $V_\mathrm B$ illustrating the ratio of couplings between the two gates to the QD formed in the middle segment and resonances attributed to QDs below the barrier gates. The couplings were characterized for an number of devices and the average values were used for compensation when acquiring the bias spectroscopy in Fig.~4\textbf{b} of the main text and for measuring $G(V_\mathrm M)$ for the QD ensemble: $V_\mathrm{T/B} = V_\mathrm{T/B}^0 + C_\mathrm{T/B,M} V_\mathrm{M}$ where $V_\mathrm{T},V_\mathrm{B} = 140, 40 \mathrm{mV}$ set the overall tuning (e.g.\ red mark in Fig.\ 4\textbf{c} the main manuscript) and the second term compensates for capacitive couplings $C_\mathrm{T,M} = -0.032 $ and $C_\mathrm{B,M} = -0.022$ between the middle and top(bottom) gates. Figure \ref{Supp_coupl}d shows a high-resolution bias spectroscopy. An example of a discrete excited state is indicated with $\Delta E \sim 0.1 E_\mathrm C$. This behavior is typical for all the acquired bias spectroscopy, and evolution of the addition energy with magnetic field was also consistent with an excited state spectrum with typical spacings of $10/20\%$ of $E_\mathrm C$.

\clearpage 
\section{Source-drain Bias Spectroscopy on Other QD Devices}

\begin{figure}[htb]
  \centering
    \includegraphics[width = \linewidth]{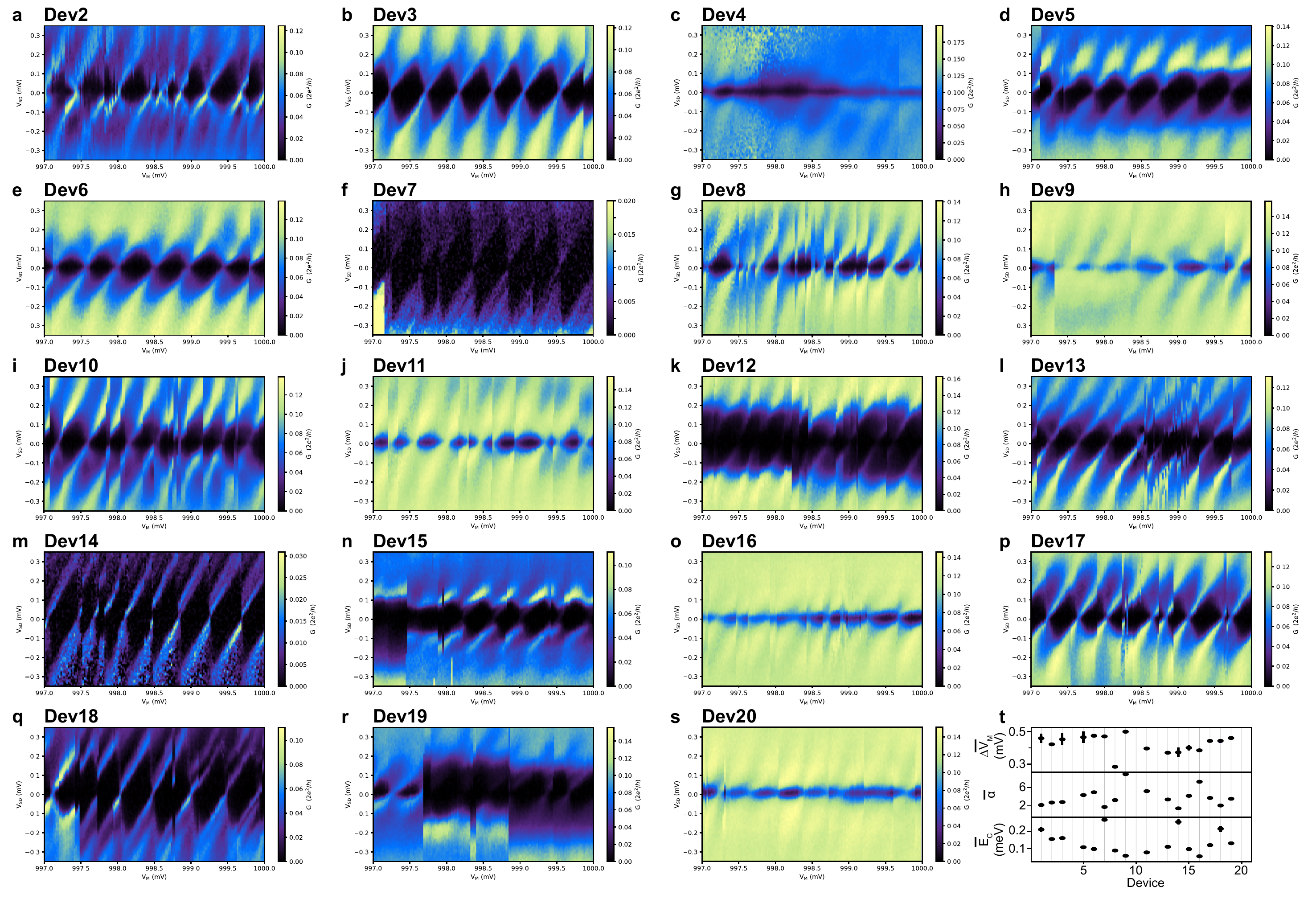}
    \caption{\textbf{a-s} Source-drain bias spectroscopy of other devices used for the QD analysis in main text. \textbf{t} Mean CB peak spacing $\overline{ V_\mathrm{M}}$, mean charging energy $\overline{ E_\mathrm{C} }$, and mean leverarm $\overline{ \alpha }$ extracted for each device with at least one full well-defined Coulomb diamond in the measurements in a-s.
    } 
    \label{Supp_bias}
  \hfill
\end{figure}

Fig.~\ref{Supp_bias}a-s shows the source-drain bias spectra of the remaining 19 devices used for the analysis. All bias spectra were taken sequentially at the $V_\mathrm{T}$, $V_\mathrm{B}$ settings as used in Fig.~4c of the main text over the same V\textsubscript{SD} and V\textsubscript{M} range. All devices exhibit Coulomb blockade physics. However, these data were taken before the results of main text Fig.~4h were compiled, and therefore not all devices were in deep Coulomb blockade at this particular gate configuration, which was chosen to optimise the behaviour of Dev1. Manually performing the analysis in main text Fig.~4h required contributions from half of the authors and took many days to complete; combining multiplexing with rapid machine-assisted analysis is the next step in simultaneously optimising device behaviour in quantum dot arrays.

The height and width of each full and well-defined Coulomb diamond was manually measured to extract the mean charging energy $\overline{ E_\mathrm{C} }$ and mean Coulomb peak spacing $\overline{ V_\mathrm{M} }$ as well as the mean leverarm $\overline{ \alpha } = \frac{\overline{V_\mathrm{M}}}{\overline{E_\mathrm{C}}}$ for each device. The resulting values are shown in Fig.~\ref{Supp_bias}t and the average of $\overline{E_\mathrm{C}}$ was used as an estimate of the QD charging energy in the analysis in the main text.

\clearpage 
\section{Examples of conductance as a function of $V_\mathrm M$}

\begin{figure}[htb]
  \centering
    \includegraphics[width = \linewidth]{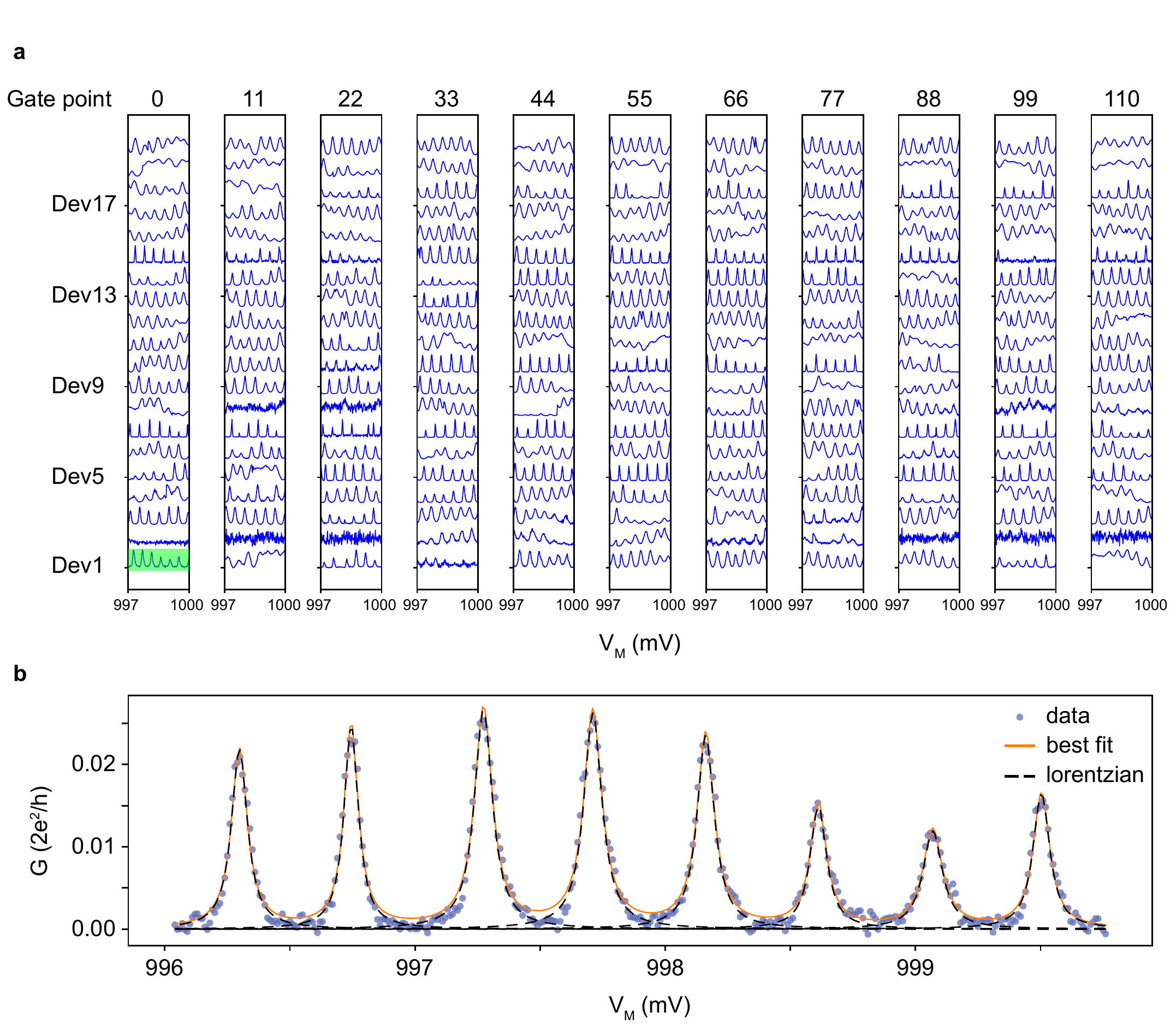}
    \caption{\textbf{a} Examples of $G(V_\mathrm M)$ normalized 0 to 1 in conductance and offset by 1.2 for gate points corresponding to the rightmost column of main text Fig. 4h ($V_\mathrm{T} = -50\, \mathrm{mV}$) \textbf{b} The raw data (blue circles) of Dev1 with $V_\mathrm{T} = -50\, \mathrm{mV}$, $V_\mathrm{B} = -50\, \mathrm{mV}$ (green box in a) with best fit (orange line) and individual lorentzians (black dashed lines).
    } 
    \label{Sup_rightmost}
  \hfill
\end{figure}

Figure \ref{Sup_rightmost}a shows examples of $G(V\mathrm M)$ normalized 0 to 1 for all devices with $V_\mathrm{T} = -50\, \mathrm{mV}$, $V_\mathrm{B} = -50\, \mathrm{mV}$ to $-200\, \mathrm{mV}$, equivalent to the rightmost column of main text Fig. 4h. The data and fit for Dev1 with $V_\mathrm{T} = -50\, \mathrm{mV}$, $V_\mathrm{B} = -50\, \mathrm{mV}$ is shown in Fig. \ref{Sup_rightmost}b.

\clearpage 
\section{Dev1 $G(V_\mathrm M)$ for all $V_\mathrm T$ and $V_\mathrm T$}

\begin{figure}[htb]
  \centering
    \includegraphics[width = \linewidth]{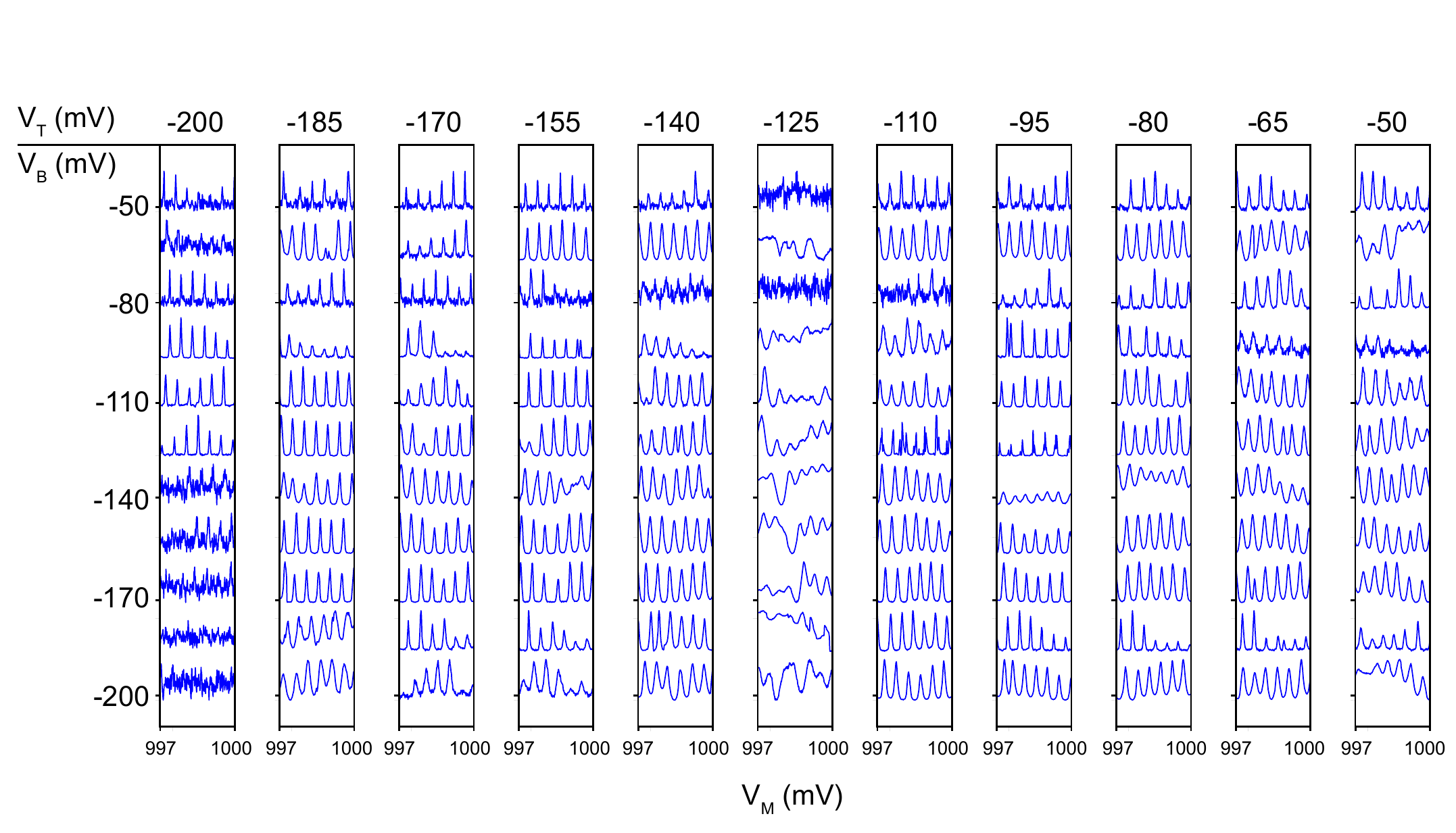}
    \caption{$G(V\mathrm M)$ normalized 0 to 1 in conductance and offset by 1.2 for all $V_\mathrm{T}$ and $V_\mathrm{B}$ for Dev1.
    } 
    \label{Sup_dev1}
  \hfill
\end{figure}

The data presented in Fig. \ref{Sup_dev1} shows $G(V\mathrm M)$ for all $V_\mathrm{T}$ and $V_\mathrm{B}$ for Dev1 normalized 0 to 1 in conductance and offset by 1.2 for clarity. The non-normalized data was used for $\Delta V_\mathrm M$ shown in main text Fig. 4d

\clearpage 
\section{Historgrams of $\Gamma$ and $\overline{\Delta V_\mathrm M}$ for Individual Devices}

\begin{figure}[htb]
  \centering
    \includegraphics[width = \linewidth]{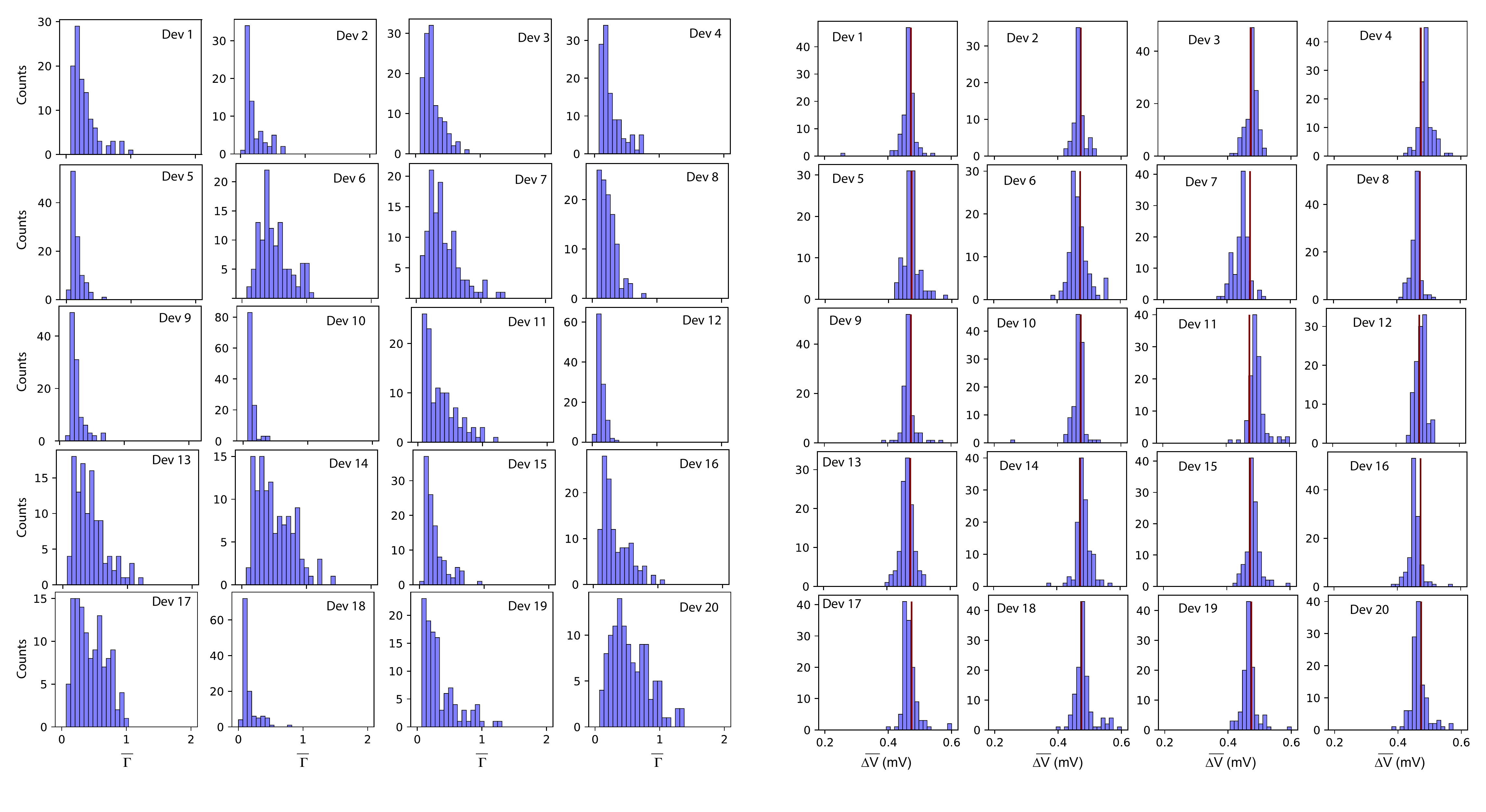}
    \caption{Histograms of $\overline{\Gamma}$ and $\overline{\Delta V_\mathrm M}$ for the individual devices The line in the $\overline{\Delta V_\mathrm M}$ histograms shows the overall mean across all devices.
    } 
    \label{FIG:individual_histograms}
  \hfill
\end{figure}

Figure \ref{FIG:individual_histograms} shows histograms of mean tunnel coupling $\overline{\Gamma}$ and Coulomb peak spacing $\overline{ V_\mathrm M }$ for each individual QD device illustrating the overall device to device variance. The red line shows the overall mean across all devices.

\bibliographystyle{naturemag}
\bibliography{ref,TSJZotero}